\documentclass[12pt]{article}
\usepackage{amsmath,amssymb,bbm,stmaryrd,graphicx}

\def\slashchar#1{\setbox0=\hbox{$#1$}           
   \dimen0=\wd0                                 
   \setbox1=\hbox{/} \dimen1=\wd1               
   \ifdim\dimen0>\dimen1                        
      \rlap{\hbox to \dimen0{\hfil/\hfil}}      
      #1                                        
   \else                                        
      \rlap{\hbox to \dimen1{\hfil$#1$\hfil}}   
      /                                         
   \fi}

\def\ad{\text{ad}}

\def\tr{\text{tr}}

\def\writecenter#1{
   \rlap{\hbox to 50mm{\hfil#1\hfil}}   
   }

\def\nn{\nonumber}

\def\be{\begin{equation}}
\def\ee{\end{equation}}
\def\ben{\begin{displaymath}}
\def\een{\end{displaymath}}
\def\bea{\begin{eqnarray}}
\def\eea{\end{eqnarray}}

\def\ft#1#2{{\textstyle {\frac{#1}{#2}} }}

\makeatletter \@addtoreset{equation}{section} \makeatother

\def\cA{A}
\def\cB{B}
\def\cC{C}
\def\cD{{\cal D}}
\def\cF{{\cal F}}
\def\cG{{\cal G}}
\def\cH{{\cal H}}
\def\cL{{\cal L}}
\def\cM{M}
\def\cO{{\cal O}}
\def\cP{{\cal P}}
\def\cQ{{\cal Q}}
\def\cU{{\cal U}}
\def\cV{{\cal V}}
\def\a{\alpha}
\def\b{\beta}
\def\c{\gamma}
\def\C{\Gamma}
\def\d{\delta}
\def\ad{{\dot\alpha}}
\def\bd{{\dot\beta}}

\def\dd{{\dot\delta}}
\def\adot{{\dot a}}
\def\bdot{{\dot b}}
\def\unA{{\underline A}}
\def\unB{{\underline B}}
\def\unC{{\underline C}}
\def\x2{X}
\def\bA{{\bar A}}

\newcommand{\w}[1]{\\[0.#1cm]}

\def\eq#1{(\ref{#1})}
\def\ft#1#2{{\textstyle{{\scriptstyle #1}\over {\scriptstyle #2}}}}

\def\s2{{\sqrt 2}}
\def\tb{{\bar\theta}}
\def\ab{{\bar A}}

\begin{document}
\begin{titlepage}
\begin{flushright}
\small UG-07-10\\
\small ENSL-00202330\\
\small MIFP-07-33\\
\end{flushright}

%
\vskip 10mm
\begin{center}
{\large {\bf THE GAUGINGS OF MAXIMAL D$=$6 SUPERGRAVITY}}
\end{center}
\vskip 8mm


\centerline{{\large \bf E. Bergshoeff $^1$, H. Samtleben $^2$
and E. Sezgin $^3$}}


\bigskip\bigskip

\begin{center}
\noindent {\it $^1$}Institute for Theoretical Physics, Nijenborgh 4,
9747 AG Groningen, The Netherlands

\medskip

\noindent {\it $^2$}Universit\'e de Lyon, Laboratoire de Physique, Ecole Normale Sup\'erieure de Lyon, \\
46 all\'ee d'Italie, F-69364 Lyon CEDEX 07, \phantom{x}France

\medskip

\noindent {\it $^3$}George P. and Cynthia W. Mitchell Institute for Fundamental
Physics,\\
Texas A\&M \phantom{x}University, College Station, TX 77843-4242, U.S.A
\end{center}

\vskip .4in

\begin{center} {\bf Abstract } \end{center}
\begin{quotation}\noindent

We construct the most general gaugings of the maximal $D=6$ supergravity. The theory is $(2,2)$ supersymmetric, and possesses an on-shell ${\rm SO}(5,5)$ duality symmetry which plays a key role in determining its couplings. The field content includes $16$ vector fields that carry a chiral spinor representation of the duality group. We utilize the embedding tensor method which determines the appropriate combinations of these vectors that participate in gauging of a suitable subgroup of ${\rm SO}(5,5)$. The construction also introduces the magnetic duals of the $5$ two-form potentials and $16$ vector fields.

\end{quotation}
\end{titlepage}

\eject


\newpage

\section{Introduction}

Gaugings of maximal supergravity theories have revealed intriguing insights into the structure of supergravity theories as well as into their higher dimensional origin and the possible symmetry structures underlying string and M-theory. The coupling of vector fields to charges assigned to the elementary fields renders the gauge theories generically non-abelian and --- more general --- in higher dimensions induces a deformation of the hierarchy of formerly abelian $p$-form tensor gauge transformations. The most systematic approach for a classification and construction of gauged supergravities resorts to exploiting the duality symmetry underlying the ungauged theories. Their possible deformations are described in terms of a constant tensor $\Theta$ encoding the embedding of the gauge group into the duality group ${\rm G}$ of the ungauged theory~\cite{Nicolai:2000sc,deWit:2002vt,deWit:2005hv}. Transforming in a certain representation of the duality group, this tensor parametrizes the possible gaugings in a manifestly covariant way. In particular, consistency of the theory can then be encoded in a number of
representation constraints on $\Theta$. The action of the gauged supergravities can be entirely parametrized by the embedding tensor; in particular, the scalar potential that arises upon gauging is given by a covariant expression bilinear in $\Theta$ dressed with the scalar fields.

From a higher-dimensional perspective a large part of the gaugings constructed in a given dimension finds a natural interpretation as the effective theories arising from compactification on curved manifolds, and/or in the presence of (geometrical and non-geometrical) fluxes (see, e.g. \cite{Grana:2005jc,Hull:2004in,Bergshoeff:2002nv}). The various geometrical and flux-parameters may be associated with the different components of the tensor $\Theta$. Vice versa, decomposing $\Theta$ under suitable subgroups of ${\rm G}$ allows to identify by merely group-theoretical methods the effective theories descending from particular compactifications. The covariant formulation of gauged supergravities furthermore allows to directly identify the transformation of the various flux parameters under the action of the duality group.

For the set of antisymmetric $p$-form tensor fields, the covariant construction of the gaugings induces a deformation of the hierarchy of formerly abelian gauge transformations. In particular, it gives rise to a St\"uckelberg-type coupling that shifts the $p$-forms with the gauge parameter of the $(p\!+\!1)$-forms. The tensor required for such a coupling that intertwines between $p$-forms and $(p\!+\!1)$-forms is proportional to the embedding tensor $\Theta$. As a consequence, the gauging non-trivially entangles the tensor gauge transformations of forms of different degree. On the level of the Lagrangian, this entanglement has an interesting consequence: while in the abelian theory all bosonic degrees of freedom are carried by the metric and antisymmetric $p$-forms with $p\le [D/2]-1$ (recall that in $D$ dimensions all higher-rank massless $p$-forms may be dualized down into massless $(D\!-\!p\!-\!2)$--forms), the generic gauging in its covariant formulation also requires explicit couplings of the $[D/2]$--forms in the action. Consistency requires that these additional forms arise with no kinetic but only a topological term (proportional to the gauge coupling constant), such that they do not introduce new propagating degrees of freedom. However, as a consequence, gauge-fixing part of the tensor gauge freedom may shuffle some degrees of freedom from the lower degree forms to the new forms, in particular render some of the latter massive. It is the specific form of the embedding tensor together with the choice of gauge fixing which encode the proper distribution of the degrees of freedom among the $p$-forms. This fits nicely with the observations in explicit compactification scenarios where turning on fluxes may induce massive $[D/2]$--forms, absent in the ungauged theory.

In even dimensions $D=2n$, there is an additional subtlety related to the fact that the duality group ${\rm G}$ of the ungauged theory is not realized off-shell but only on the combination of equations of motion and Bianchi identities of the $(n\!-\!1)$--forms. More specifically, only (the ``electric'') half of the $(n\!-\!1)$--forms shows up in the Lagrangian while the other half is defined as their on-shell (``magnetic'') duals. Only together they form an (irreducible) representation of ${\rm G}$. Upon gauging, both electric and magnetic $(n\!-\!1)$--forms enter the Lagrangian; again the latter couple only with a topological term in order to preserve the balance of degrees of freedom. Contrary to what one might expect at first glance, the construction allows even for the gauging of subgroups of ${\rm G}$ that are not off-shell realized in the ungauged theory. In other words, there is a well-defined Lagrangian even for such gaugings whose gauge group is not among the global symmetries of the ungauged Lagrangian. The existence of these gaugings is intimately related to the appearance of magnetic forms in the action. This construction has been worked out in 4 dimensions~\cite{deWit:2005ub,de Wit:2007mt} where the relevant duality is electric/magnetic duality for vector fields and in 2 dimensions~\cite{Samtleben:2007an} where it amounts to the scalar-scalar duality which is at the heart of the integrable structure of the ungauged theory.

In this paper we consider the maximal $D=6$ supergravity and its possible gaugings. The ungauged maximal supergravity in six dimensions has been constructed in~\cite{Tanii:1984zk} and possesses a global ${\rm E}_{5(5)}={\rm SO}(5,5)$ symmetry. Only a ${\rm GL}(5)$ subgroup is realized off-shell with the $5$ two-forms $B_{m}$ transforming in its fundamental representation. Together with their magnetic duals $B^{m}$ in the $5'$ they combine into the vector representation ${\bf 10}$ of ${\rm SO}(5,5)$. Little is known about the gaugings of this theory. Cowdall~\cite{Cowdall:1998rs} obtained an ${\rm SO}(5)$ gauge theory from circle reduction of the ${\rm SO}(5)$ gauged maximal supergravity in 7D \cite{Pernici:1984xx}. Alternatively, this theory describes the $S^4$ reduction of the IIA theory and proves to be relevant in a non-conformal extension of the AdS/CFT correspondence~\cite{Boonstra:1998mp}. However, as it has only the ${\rm SL}(5)$ symmetry inherited from $7D$ manifest, the $6D$ result is in an exceedingly complicated form that does not shed much light onto the maximal duality symmetry. Here we fill this gap by providing all possible gaugings by a direct construction in $6D$. The embedding tensor $\Theta$ which covariantly parametrizes the possible deformations transforms in the ${\bf 144}_c$ spinorial representation of ${\rm SO}(5,5)$. The gauged Lagrangian features the full set of ${\bf 10}$ two-forms as well as a set of three-forms in the ${\bf 16}_{s}$ which are on-shell dual to the vector fields of the theory. We should stress that our formalism differs from other approaches introducing $p$-form fields together with their duals in that the relevant first order duality equations here arise as true equations of motion from the Lagrangian. This appears only possible in the gauged theory.

The plan of this paper is the following. In section~\ref{sec:ungauged} we review the building blocks of maximal $D=6$ supergravity. In particular, we discuss the role of the ${\rm SO}(5,5)$ duality group under which electric and magnetic two-forms undergo an orthogonal rotation and their consistent coupling is provided by the formalism of Gaillard and Zumino \cite{Gaillard:1981rj}. We review in detail the structure of the scalar fields which parametrize the coset space ${\rm SO}(5,5)/({\rm SO}(5)\times{\rm SO}(5))$. Finally, we give Tanii's Lagrangian of the ungauged theory. In section~\ref{sec:gauged} we turn to the gauging of the theory. Applying the general framework, the gauging is parametrized by the embedding tensor $\Theta$ transforming in the ${\bf 144}_c$ of ${\rm SO}(5,5)$. We derive the quadratic constraints on this tensor whose solutions correspond to viable gaugings of the six-dimensional theory and work out the deformed tensor hierarchy up to and including the three-forms. We present the Lagrangian of maximal gauged $D=6$ supergravity which for a general gauging carries the set of 10 electric and magnetic two-forms  $B_{M}=(B_{m}, B^{m})$ of which the latter couple only with a topological term $\Theta\,C\,dB$ to the set of three-forms $C_{A}$ in the ${\bf 16}_{s}$. Finally, we give a short overview and discussion of various types of possible gaugings, i.e.\ solutions of the quadratic constraint and discuss their possible higher-dimensional origin by dimensional reduction from seven and eleven dimensions, respectively. Furthermore, we discuss the truncation to $N=(1,1)$ theories. Our notations and conventions are given in Appendix A, and some identities, useful in deriving the topological Lagrangian and computing its variation, are given in Appendix B.

\section{The Ingredients of the Maximal D$=$6 Supergravity}\label{sec:ungauged}
\subsection{The Field Content}

The $N=(2,2)$ supersymmetric maximal supergravity in six dimensions has been constructed by Tanii \cite{Tanii:1984zk}. It is an ungauged theory in which the couplings are governed, along with supersymmetry, by the duality symmetry group ${\rm SO}(5,5)$ that rotate the field equations and Bianchi identities of the five 2-form potentials into each other. Only the subgroup ${\rm GL}(5)\subset {\rm SO}(5,5)$ is a manifest off-shell symmetry of the theory. There is also a manifest composite local symmetry ${\rm SO}(5) \times {\rm SO}(5)$.

The bosonic fields of the theory are the vielbein $e_\mu^r$,  2-form potentials $B_{\mu\nu m} (m=1,...,5)$, vector fields $A_\mu^\cA (\cA=1,...,16)$ and scalars $V_\cA^{\a\ad} (\a,\ad=1,...,4)$ that parametrize the coset ${\rm SO}(5,5)/({\rm SO}(5)\times {\rm SO}(5))$. The index $\cA$ labels the $16$ dimensional Majorana-Weyl spinor of ${\rm SO}(5,5)$, and the indices $\a,\ad$ label the spinors of ${\rm SO}(5)\times {\rm SO}(5)$. The spinor fields are the gravitini $\psi_{+\mu\a}$, $\psi_{-\mu\ad}$ and $\chi_{+a\ad}$, $\chi_{-\adot\a}$, where $a,\adot=1,...,5$ are the ${\rm SO}(5)\times {\rm SO}(5)$ vector indices, and $\pm$ refers to the spacetime chirality of the spinors which are symplectic-Majorana-Weyl. (See Appendix A for further notations and conventions). In summary, the full supergravity multiplet consists of the fields:
\be
\left(\ e_\mu^r,\ B_{\mu\nu m},\ \cA_\mu^\cA,\ V_A^{\a\ad},\ \psi_{+\mu \a},\ \psi_{-\mu \ad},\ \chi_{+a\ad},\ \chi_{-\adot\a}\ \right)\ .
\ee
As we gauge this theory in the most general possible way, we will introduce the following duals of the vector fields and the $2$-form potentials:
\be
\left(\ B_{\mu\nu}{}^m,\ C_{\mu\nu\rho \cA}\ \right)\ .
\ee
Note that the vectors are in ${\bf 16_c}$ and the $3$-form potentials in ${\bf 16}_s$ of the duality group ${\rm SO}(5,5)$. Electric and magnetic two-forms $B_m$ and $B^m$ transform in the $5$ and $5'$ of ${\rm GL}(5)$, respectively, and combine into the ${\bf 10}$ of ${\rm SO}(5,5)$.

From $E_{11}$, it has been predicted that one can extend the field content
of $D=6$ maximal gauged supergravity by the introduction of further $4,5$ and $6$-forms \cite{Riccioni:2007au,Bergshoeff:2007qi}:
\be
\left(\,C^{(4)}_{MN},\ \ C^{(5)}_{M\cA},\ \ C^{(6)}_{MN,P}\ ,\ \ C^{(6)}_{MNPQR+}\,\right) \ ,
\label{hpf}
\ee
where $C_{MN}$ is antisymmetric, $C_{M\cA}$ is $\c$-traceless, $C_{MN,P}$ is mixed symmetric, $C_{MNPQR+}$ is self-dual, and thus in ${\bf 45}$, ${\bf 144_s}$, ${\bf 320} +{\bf 10}$ and ${\bf 126_s}$ dimensional representations of ${\rm SO}(5,5)$, respectively. The $4$-form potentials have constraints on their curvatures such that on-shell they describe $25$ independent degrees of freedom corresponding to the Hodge duals of the scalar fields in the coset ${\rm SO}(5,5)/{\rm SO}(5)\times {\rm SO}(5)$. We will see that the $5$-forms are in the same representation as the embedding tensor and that the quadratic constraints of the embedding tensor precisely transform in the representations dual to the $6$-forms given in \eq{hpf} \cite{Bergshoeff:2007vb,dWS,dWS2}. These 5-forms and $6$-forms can easily be included in the $D=6$ Lagrangian, where the constant embedding tensor has been replaced by a scalar field, as Lagrange multipliers giving rise to the constancy of the embedding tensor and the quadratic constraints, respectively \cite{Bergshoeff:2007vb,dWS,dWS2}. We will not explicitly perform this construction in this paper. Recently, $D=5$ maximal gauged supergravity has been constructed using the embedding tensor approach and its relation with an $E_{11}$-extended spacetime has been investigated \cite{Riccioni:2007ni}. It would be interesting to further study the proposed relationship for the six-dimensional case studied in this paper.

\subsection{Duality Symmetry}

To appreciate the duality symmetries in Tanii's Lagrangian and also to set our notation, we begin by reviewing the part of the Lagrangian involving the 2-form potential. Let us define the field strengths

\be H^{(0)}_m = dB_m \ , \qquad
 {\widetilde G}_{(0)}^m = -3! e^{-1}{\partial {\cal L}\over \partial H^{(0)}_m}\ .\label{fg}
\ee
The Hodge-dual of a 3-form $\omega$ is defined as $\tilde\omega_{\mu\nu\rho}= \ft1{3!} e\epsilon_{\mu\nu\rho\sigma\kappa\lambda}\, \omega^{\sigma\kappa\lambda}$. The field equations $dG_{(0)}^m=0$ and the Bianchi identities $dH^{(0)}_m=0$ form a system invariant under linear transformations, which are restricted to ${\rm SO}(5,5)$ by the requirement that the equation for $G_{(0)}^m$ is covariant under these transformations. Infinitesimally, these transformations act as
\be
\delta \left(\begin{array}{c}H^{(0)}_m\\G_{(0)}^m\\ \end{array}\right) = u
\left(\begin{array}{c}H^{(0)}_m\\G_{(0)}^m \\ \end{array}\right)\ ,
\qquad u^T\eta +u\eta=0\ ,
\qquad \eta=\left(\begin{array}{cc}0 & \delta_m^n\\ \delta_n^m & 0\\ \end{array} \right)\ .
\label{eta}
\ee
Gaillard and Zumino have shown that the appropriate Lagrangian that achieves the duality symmetry is given by \cite{Gaillard:1981rj}\footnote{For a very nice review, see \cite{Tanii:1998px}.}
\be \cL= -\ft1{12} e H^{(0)}_m {\widetilde G}_{(0)}^m -\ft1{12}e \left[H^{(0)}_m S^m + G^{(0)m} R_m \right]
+\cL_{inv}\ , \label{gz}
\ee
where $(R_m, S^m)$, which is a pair that transforms under ${\rm SO}(5,5)$ as in \eq{eta}, and $\cL_{inv}$, which is duality invariant, are built out of fields other than $(H_m^{(0)}, G_{(0)}^m)$, and $jG^m_{(0)}$ is given by
\be
jG^m_{(0)}= S^m+K^{mn}(H^{(0)}_n-jR_n)\ .\label{jg}
\ee
The operation $j$ acting on a given 3-form $\omega$ is defined by
\be j\omega=\tilde \omega\ ,  \qquad j^2=+1\ ,\qquad \tilde\omega_{\mu\nu\rho}= \ft1{3!} e\epsilon_{\mu\nu\rho\sigma\kappa\lambda}\, \omega^{\sigma\kappa\lambda}\ ,
\ee
and the matrix $K^{mn}$ to be built out of the scalar fields must be of the form
\be
K^{mn}= K_+^{mn}P_+ + K_-^{mn}P_-\ ,  \qquad  (K_+)^T = K_-\ ,\qquad P_\pm=\frac12 (1\pm j)\ ,
\label{kpm}
\ee
or equivalently
\be
K^{mn}= K_1^{mn} +j K_2^{mn}\ , \qquad K_1^T=K_1\ ,\qquad K_2^T=-K_2\ .
\label{kmn2}
\ee
Under the infinitesimal ${\rm SO}(5,5)$ duality transformations,
\be
u=\left(\begin{array}{cc} x & y \\z & t \\ \end{array}\right)\ ,
\label{dt1}
\ee
$K$ must transforms as
\be
\delta K= -K x +tK +zj -KyKj \ , \label{dt}\ee
as required by the covariance of the second equation in \eq{fg}. For the $5\times 5$ matrices $K_\pm$ this gives
\bea
\delta K_+ &=& -K_+ x+t K_+ +z - K_+ y K_+\ ,
\nn\w2
\delta K_- &=& -K_- x+t K_- -z + K_- y K_-\ .
\eea
Substituting \eq{jg} into the Lagrangian \eq{gz} gives
\be
 e^{-1}{\cal L}= -\ft1{12}  \left( H_m^{(0)}-jR_m\right) K^{mn} \left( H_n^{(0)}- jR_n\right)
-\ft16 \left(H_m^{(0)}-jR_m\right) \cdot S^m -\ft1{12} jR_m\cdot S^m + e^{-1} {\cal L}_{inv}\ .
\ee
%

\subsection{Gauge Symmetry}

So far the construction is rather general, and as far as duality symmetry is concerned the result above provides the answer.  In the particular model we wish to study, however, we need to consider the gauge symmetries and supersymmetry as well. To this end, we need to introduce the Chern-Simons modified 3-form field strengths, and their duality invariant Pauli couplings to fermionic bilinears. To achieve this, the pair $(R_m,S^m)$ is chosen as

\be
jR_m= -\omega_m+{\cal O}_m\ ,\quad\quad jS^m= -\omega^m +{\cal O}^m\ ,
\ee
where the Chern-Simons  forms are given by

\be \omega_m=\ft1{\s2}\, {\bar F}\wedge \gamma_m A\ ,\qquad \omega^m = \ft1{\s2}\, {\bar F}\wedge \gamma^m A \ ,
\ee
and we have used the $16\times 16$ chirally projected ${\rm SO}(5,5)$ Dirac matrices $\c_M=(\c_m, \c^m)$, and $(\cO_m,\cO^m)$
are bilinears in fermions, to be determined by supersymmetry, multiplied by suitable functions of the scalar fields so
that they transform as $(R_m,S^m)$ under ${\rm SO}(5,5)$ transformations. Thus, the Lagrangian takes the form
\bea
e^{-1} \cL &=&  -\ft1{12} H_m \cdot K^{mn} H_n +\ft16 H_m \cdot j\omega^m -\ft1{12} \omega_m \cdot j \omega^m
\nn\\
&&+\ft16 H_m \cdot \left( K^{mn}\cO_n -j\cO^m \right)-\ft1{12} \cO_m \cdot \left(K^{mn}\cO_n -j\cO^m\right) \nn\\
&& +\ft1{12} \left( \omega_m \cdot j\cO^m +\omega^m  \cdot j \cO_m \right) + e^{-1} \cL_{inv}\ ,
\eea
where
\be H_m=H^{(0)}_m + \omega_m \ .\label{fs}\ee
Given the gauge transformations
\be
\delta B_m=-\ft1{\s2}\, {\bar F}\gamma_m \lambda\ ,\qquad \delta A= d\lambda\ ,
\ee
we see that all but the $\omega \cO$ terms are invariant, since
\be {\bar F}\wedge \gamma_M F \wedge {\bar F}\gamma^M \lambda=0 \ ,\ee
which holds, thanks to the well known identity
\be  \c^{\vphantom{M}}_{M(AB}\,\c^M_{C)D} =0\ .\ee
As to the $\omega \cO$ terms, while they are not gauge invariant, they are nonetheless duality invariant. Therefore, we can discard them by choosing $\cL_{inv}$ to contain these terms with opposite sign. Then, we are left with
\bea
e^{-1} \cL &=&  -\ft1{12} H_m \cdot K^{mn} H_n +\ft16 H_m \cdot j\omega^m -\ft1{12} \omega_m \cdot j \omega^m
\nn\\
&&+\ft16 H_m \cdot \left( K^{mn}\cO_n -j\cO^m \right)-\ft1{12} \cO_m \cdot \left(K^{mn}\cO_n -j\cO^m\right) + e^{-1} \cL'_{inv}\ .\label{L1}
\eea
The Lagrangian is then determined completely by specifying $K^{mn}$, the pair of 3-forms $(\cO_m,\cO^m)$ and $\cL'_{inv}$.
Defining a following dual field strength, in analogy with \eq{fs},
\be
G^m= G^m_{(0)}+\omega^m\ ,
\ee
it follows from \eq{jg} that
\be
G^m = j K^{mn} H_n + \cO^2\ {\rm terms}\ .\label{gk}
\ee
In the supergravity model we shall study, $\cO^2$ represents quartic fermion terms. Working up to quartic fermion terms in the action, which we shall do in the rest of the paper, it is convenient to define field strengths $G_M$ that transform as $10$-plet of the duality group ${\rm SO}(5,5)$ as
\be
G_M = \left(\begin{array}{c}  G_m\\   G^m \\ \end{array}\right) =
\left(\begin{array}{c} H_m\\  j K^{mn} H_n \\ \end{array}\right)
\label{d31}
\ee
Using this definition, the Lagrangian \eq{L1} can be written as
\bea
e^{-1} \cL &=&  -\ft1{12} H_m \cdot K^{mn} H_n +\ft16 H_m \cdot j\omega^m -\ft1{12} \omega_m \cdot j \omega^m
\nn\\
&&+\ft16 jG_M \cdot \cO^M + e^{-1} \cL'_{inv}\ ,\label{L11}
\eea
where $\cO^M= (\cO^m,\cO_m)$ and we have dropped $\cO^2$ terms that are quartic in fermions. With $\cO$ representing fermionic bilinears, the $jG\cdot \cO$ term describes already duality invariant Pauli couplings.

Next, we discuss the matrix $K^{mn}$ which is to be expressed in terms of the scalar fields, following \cite{Tanii:1984zk}. Here we shall choose a convenient basis for the scalar fields to make the ${\rm GL}(5)\in {\rm SO}(5,5)$ symmetry manifest at the Lagrangian level. To this end, we introduce the $10\times 10$ matrix
\be {\cal V}_M{}^\unA= \left(
                      \begin{array}{cc}
                        \cV_m{}^a & \cV_m{}^\adot \\
                        \cV^{ma} & \cV^{m \adot} \\
                      \end{array}
                    \right) \equiv \left( \begin{array}{cc} A & B \\ C & D \\ \end{array} \right)
\label{10}
\ee
where $a,\adot$ are the vector indices of ${\rm SO}(5)\times {\rm SO}(5)$. Tanii has expressed his results in a basis in which $ H^{(0)}_m \pm G_{(0)}^m $ transform into each other under ${\rm SO}(5,5)$ as components of 10-vector, and used a matrix $\cU$ that obeys the relation
\be
\cU^T \eta_{diag}\cU=\eta_{diag}\ , \qquad  \eta_{diag}= {\rm diag}\,(1,-1)\ ,
\ee
and therefore it is an ${\rm SO}(5,5)$ representation. However, in this basis, the ${\rm GL}(5)$ symmetry is not manifest. This can be remedied by working in a basis in which $(\,H^{(0)}_m, G_{(0)}^m\,)$ transform as a vector under ${\rm SO}(5,5)$. To achieve this, we work with the matrix $\cV$ of \eq{10} which is related to the group element $\cU$ as
\be
\cV=M\cU\ ,  \qquad M={1\over \sqrt 2} \left(\begin{array}{cc}1& 1 \\ 1 & -1 \\ \end{array}\right)\ .
\ee
Since $M^T\eta_{diag} M=\eta$ with $\eta$ defined as in \eq{eta}, the matrix $\cV$ satisfies the relation
\be \cV^T \eta \cV= \eta_{diag}\ , \label{oc}    \ee
where $\eta$ is as given in \eq{eta}, and $\eta_{diag}$ explicitly by
\be
\eta_{\unA \unB}= \left( \begin{array}{cc} \delta_{ab} & 0\\ 0 & -\delta_{\adot\bdot}\\ \end{array} \right)
\ee
From \eq{oc}, it also follows that
\bea
&& \cV^{Ma} \cV_M{}^b = \delta^{ab}\ ,\qquad \cV^{M\adot} \cV_M{}^\bdot = -\delta^{\adot\bdot}\ ,\qquad
\cV^{Ma} \cV_M{}^\adot = 0\ ,
\nn\w2
&& \cV_M{}^a \cV^{Na} - \cV_M{}^\adot \cV^{N\adot} = \delta_M^N\ .
\label{oc3}
\eea
It is important to note that in our conventions, the explicitly written $(a, \adot)$ indices are always raised and lowered with $+\delta_{ab}$ and $+\delta_{\adot\bdot}$, starting from the basic object \eq{10}. This explains the occurrence of minus signs in the formulae above where the form of $\eta_{\unA\unB}$ has been used.

Our choice of the scalar matrix $\cV$ makes both the  ${\rm GL}(5)$ acting from the left, and ${\rm SO}(5)\times {\rm SO}(5)$ acting from the right manifest in the formalism. Note that, given $\cV$, the group $h={\rm SO}(5)_I\times {\rm SO}(5)_{II}$ acts from the right diagonally in the form $h= {\rm diag}\, (h_I, h_{II})$. The condition \eq{oc} translates into
\be A^T C+C^T A = 1\ ,\qquad  B^T D+D^T B=-1\ ,\qquad A^T D + C^T B =0\ . \label{oc2}\ee
With this parametrization, the matrix $K^{mn}$ can be chosen as\footnote{We are grateful to Yoshiaki Tanii for helpful discussions regarding this point.}
\be
K=CA^{-1}P_+ - DB^{-1} P_-\ .
\label{kca2}
\ee
Using \eq{oc2}, one finds that $(CA^{-1})^T=-DB^{-1}$. It can be easily checked that this $K$ indeed transforms under ${\rm SO}(5,5)$ as in \eq{dt}. Written in terms of $\cV$, we have
\be
K^{mn}= \cV^{ma}(\cV_n{}^a)^{-1} P_+ - \cV^{m\adot}(\cV_n{}^\adot)^{-1} P_-\ ,
\label{K}
\ee
which gives the useful relations
\be
K_+^{mn} \cV_n{}^a  = P_+\cV^{ma}\ ,\qquad K_-^{mn}\cV_n{}^\adot= - P_-\cV^{m\adot}\ ,
\label{kv}
\ee
with $K_\pm$ defined in \eq{kpm}.

\subsection{Supersymmetry}

The choice for $\cO_M$ is dictated by supersymmetry. Tanii has found that the following choices are appropriate \cite{Tanii:1984zk}
\be
\cO_M = \ft1{\s2}\,\left( \cV_M{}^a P_- \cO^a + \cV_M{}^\adot P_+ \cO^\adot\right)\ ,
\ee
with the exact form of the fermionic bilinears $(\cO_a, \cO_\adot)$ determined by supersymmetry (see next section). Moreover, the description of the supersymmetric transformation rules requires the quantities $H^a$ and $H^\adot$ defined by
\be
H_m = \s2\, \left(\cV_m{}^a P_+ H^a - \cV_m{}^\adot P_- H^\adot\right)\ .
\label{hm1}
\ee
Recalling \eq{kv} and \eq{oc3}, we find that\footnote{The indices $(a,\adot)$ on $H$, ${\cal O}$ and $\cV$ are raised and lowered by $\delta_{ab}$ and $\delta_{\adot\bdot}$.}
\be
P_+ H_a = \ft1{\s2}\,P_+G_M \cV^M_a\ , \qquad P_- H_\adot = -\ft1{\s2}\,P_-G_M \cV^M_\adot\ .
\label{ha1}
\ee
Employing the relations \eq{K} and \eq{oc3} also shows that
\be
P_+G_M \cV^M_\adot=0\ ,\qquad P_-G_M \cV^M_a=0\ .
\label{ha2}
\ee
 Using the quantities defined so far, the Lagrangian \eq{L11} can be written as
\bea
 e^{-1}\cL &=& -\ft1{12} H_m \cdot K^{mn} H_n +\ft16 \left( P_+ H^a  \cdot \cO^a
 +P_-H^\adot \cdot \cO^\adot \right)
\nn\\
&& +\ft16\, H_m \cdot j\omega^m -\ft1{12} \omega_m\cdot j \omega^m + e^{-1} \cL''_{inv}\ .
\label{L2}
\eea
In showing the cancelations of the terms proportional to $\psi_\mu H^2$ terms coming from the variation of the metric in the $H$-kinetic terms, it is useful to note that
\be
\delta \cL_{\rm kin}(H)= -\ft14 e K_1^{mn} \left( H_{\mu m}^+ \cdot H_{\nu n}^+ + H_{\mu m}^- \cdot H_{\nu n}^-\right) \delta g^{\mu\nu}\ ,
\ee
where $H^\pm \equiv P_\pm H$, and we have used the identity
\be
K_1^{mn} H_{\mu m}^+ H_{\nu n}^- +  (\mu \leftrightarrow \nu)  = \ft13 g_{\mu\nu}\, K_1^{mn} H_m^+ \cdot H_n^-\ ,
\ee
where we have used $K_1^{mn} \cV_m^a \cV_n^b=\ft12 \delta^{ab}$, which follows from
\eq{kmn2}, \eq{oc2} and \eq{kca2}. We are also using the notation $H_\mu \cdot H_\nu\equiv H_{\mu \rho\sigma}H_\nu{}^{\rho\sigma}$, and $H^+\cdot H^- \equiv H_{\mu\nu\rho}^+ H^{-\mu\nu\rho}$. Finally, upon using \eq{hm1} and \eq{oc3} one finds that
\be
\delta \cL_{\rm kin}(H)= -\ft14 e \left(  H_\mu^{+a} \cdot H_\nu^{+a} +  H_\mu^{-\adot} \cdot H_\nu^{-\adot}\right) \delta g^{\mu\nu}\ .
\ee
These terms are then canceled by terms arising from the variation of the Pauli couplings in \eq{L2}.

\subsection{The Scalars}

The $25$ scalar field of the theory parametrize the coset ${\rm SO}(5,5)/({\rm SO}(5)\times {\rm SO}(5))$ which can conveniently be parametrized in terms of an ${\rm SO}(5,5)$ valued $16\times16$ matrix $V_\cA{}^{\a\bd}$, with its inverse defined by
\be
V_\cA{}^{\a\bd}\,V^\cB{}_{\a\bd} = \delta_\cA^\cB\ ,\qquad
V_\cA{}^{\a\bd}\, V^\cA{}_{\c\dd} ~=~ \delta^\a_\c\,\delta_\dd^\bd\ .
\ee
The $10\times 10$ scalar matrix $\cV$ defined in \eq{10} can be expressed in terms of the above $16\times 16$ matrices $V$ as \footnote{Working with $\cV$ related to ${\rm SO}(5,5)$ matrices $\cU$ through $\cV= M\cU$ implies that the ${\rm SO}(5,5)$ $\c$ matrices obey the Clifford algebra with off diagonal $\eta_{MN}$.} (see Appendix A for notations and conventions).
\be
\cV_M{}^a = \ft1{16} {\bar V}\gamma_M\gamma^a V\ , \qquad  \cV_M{}^\adot = -\ft1{16} {\bar V}\gamma_M\gamma^\adot V\ . \label{gr1}
\ee
These relations follow from the fact that the ${\rm SO}(5,5)$ $\c$-matrices (obeying Clifford algebra with non-diagonal $\eta_{MN}$) are left invariant by ${\rm SO}(5,5)$ transformations realized in terms of $V$ and ${\cV}$. Noting that (see Appendix A)
\be
\cV_M{}^\unA \C_\unA = \left(
                         \begin{array}{cc}
                           0 & \cV_M{}^a \c^a + \cV_M{}^\adot \c^\adot \\
                          \cV_M{}^a \c^a - \cV_M{}^\adot \c^\adot & 0  \\
                         \end{array}
                       \right)\ ,
\ee
the invariance of the ${\rm SO}(5,5)$ $\c$-matrices translates into the relations
\bea
V_{A\a\ad} (\c_M)
^{AB} V_B^{\b\bd}
&=& \cV_M{}^a\, (\c^a)_\a{}^\b \delta_\ad^\bd +\cV_M{}^\adot (\c^\adot)_\ad{}^\bd \delta_\a^\b\ ,
\nn\w2
V^A_{\a\ad} (\c_M)_{AB} V^{B\b\bd}
&=& \cV_M{}^a\, (\c^a)_\a{}^\b \delta_\ad^\bd -\cV_M{}^\adot (\c^\adot)_\ad{}^\bd \delta_\a^\b\ ,
\label{gr2}
\eea
from which \eq{gr1} follows.

The scalar currents are defined as \cite{Tanii:1984zk}
\be
V^\cA{}_{\a\ad}\; \partial_\mu V_\cA{}^{\b\bd}
=
\ft14 Q_\mu^{ab}\, (\c^{ab})_\a{}^\b \delta_\ad^\bd + \ft14 Q_\mu^{\adot\bdot}\, \delta_\a^\b (\c^{\adot\bdot})_\ad{}^\bd + \ft14 P_\mu^{a\adot}\, (\c^a)_\a{}^\b (\c^\adot)_\ad{}^\bd\ .
\ee
It follows that
\be
P_\mu^{a\adot} = \ft14 {\bar V} \c^a\c^\adot \partial_\mu V\ ,\qquad Q_\mu^{ab} = \ft18 {\bar V} \c^{ab} \partial_\mu V\ ,
\qquad Q_\mu^{\adot\bdot} = \ft18 {\bar V} \c^{\adot\bdot}\partial_\mu V\ ,
\ee
and
\be
D_\mu V = \ft14\, P_\mu^{a\adot}\, \c^a\c^\adot V\ .
\ee
Moreover, we have the standard integrability conditions
\be
D_{[\mu}P_{\nu]}^{a\adot}=0\ ,\quad \partial_{[\mu}Q_{\nu]}^{ab} +Q_{[\mu}{}^{ac} Q_{\nu]}{}^{cb}+ \ft14 P_{[\mu}^{a\adot}P_{\nu]}{}^{\adot b}=0 \ , \label{2}
\ee
and a similar equation for the curl of $Q_\mu^{\adot\bdot}$. The covariant derivatives in the above expressions contain the composite connections. Other useful identities are:
\be
D_\mu \cV_M{}^a= \ft12\, P_\mu^{a\adot}\,\cV_M{}^\adot\ , \qquad
D_\mu \cV_M{}^\adot = \ft12\, P_\mu^{a\adot}\,\cV_M{}^a\ .  \label{3}
\ee
It is also useful to introduce the matrix
\be
\cM_{\cA\cB}= V_\cA{}^{\a\bd} V_{\cB\a\bd}\ ,
\ee
which will be used in the construction of kinetic term for the vector fields.

\subsection{The Lagrangian}\label{sec:Tanii}

Using the building blocks describe above, Tanii's Lagrangian \cite{Tanii:1984zk}, can be written in
our notation and conventions (see Appendix A) as follows:
\be
\cL=\cL_B+\cL_F\ ,\label{LT}
\ee
where
\bea
e^{-1} \cL_B &=&  \ft14 R  -\ft1{12} H_m \cdot K^{mn} H_n -\ft14 \cM_{AB} F_{\mu\nu}^\cA F^{\mu\nu\cB}-\ft1{16} P_\mu^{a\adot} P^\mu_{a\adot}
\nn\\
&&+\ft16\, H_m \cdot j\omega^m -\ft1{12}\omega_m\cdot j \omega^m \ ,
\label{LB}
\eea
and, up to quartic fermions,\footnote{We have corrected the coefficient of the $P_\mu I^\mu$ term in \cite{Tanii:1984zk}.}
\bea
e^{-1} \cL_F &=& -\ft12 {\bar\psi}_{+\mu} \c^{\mu\nu\rho} D_\nu \psi_{+\rho}
-\ft12 {\bar\psi}_{-\mu} \c^{\mu\nu\rho} D_\nu \psi_{-\rho}
-\ft12 {\bar\chi}^a \c^\mu D_\mu \chi^a
-\ft12 {\bar\chi}^\adot\c^\mu D_\mu \chi^\adot \nn\w2
&&
+\ft1{4}P_\mu^{a\adot}\,I^\mu_{a\adot} - \ft12 F_{\mu\nu}^\cA\,J^{\mu\nu}_\cA
+\ft16 \left( P_+H^a \cdot \cO^a + P_-H^\adot \cdot \cO^\adot \right) \ .
\label{LF}
\eea
The fermionic bilinears occurring in \eq{LF} have been determined by Tanii as follows
\bea
I_\mu^{a\adot}&=&
{\bar\psi}_\nu \c^\mu\c^\nu \c^a \chi^\adot + {\bar\psi}_\nu\ \c^\mu\c^\nu \c^\adot \chi^a\ ,
\nn\w2
J_{\mu\nu\cA} &=&
{\bar\psi}_\rho\c^{[\rho}\c_{\mu\nu}\c^{\sigma]} V_\cA \psi_\sigma
+\ft12 {\bar\psi}_\rho\c_{\mu\nu}\c^\rho\c^a V_\cA \chi^a
\nn\w2
&& -\ft12 {\bar\chi}^\adot\c^\rho\c_{\mu\nu} V_\cA\c^\adot\psi_\rho
+\ft12 {\bar\chi}^\adot \c_{\mu\nu}\c^a V_\cA\c^\adot \chi^a\ ,
\nn\w2
\cO_{\mu\nu\rho}^a &=&
3{\bar\psi}_{[\mu} \c_\nu\c^a \psi_{\rho]}
-3 {\bar\psi}_{[\mu}\c_{\nu\rho]} \chi^a
-\ft14 {\bar\chi}^\bdot \c_{\mu\nu\rho}\c^a\chi^\bdot\ .
\label{fb}
\eea
and $\cO^\adot$ is obtained from $\cO^a$ by interchanging dotted and undotted indices.

The Lagrangian \eq{LT} is invariant under the following supersymmetry transformations:
\bea
\delta e_\mu{}^r &=& {\bar\epsilon}_+\c^r \psi_{+\mu}+ {\bar\epsilon}_-\c^r\psi_{-\mu}\ ,
\nn\w2
\delta\psi_{\mu +}&=& D_\mu\epsilon_+ -\ft1{24} H_{\rho\sigma\kappa}^a\c^a\c^{\rho\sigma\kappa}\c_\mu\epsilon_+
+\ft18 \left(\c_\mu{}^{\nu\rho}-6\d_\mu^\nu\c^\rho\right) F_{\nu\rho}^\cA V_\cA\, \epsilon_-\ ,
\nn\w2
\delta\psi_{\mu -}&=& D_\mu \epsilon_- -\ft1{24} H_{\rho\sigma\kappa}^\adot\c^\adot\c^{\rho\sigma\kappa}\c_\mu\epsilon_-
+\ft18 \left(\c_\mu{}^{\nu\rho}-6\d_\mu^\nu\c^\rho\right) F_{\nu\rho}^\cA {\tilde V}_A\, \epsilon_+\ ,
\nn\w2
\delta\chi^\adot &=& \ft14 P^{\mu a\adot} \c^a \c^\mu \epsilon +\ft1{12} H_{\mu\nu\rho}^\adot \c^{\mu\nu\rho}\epsilon +\ft18 F_{\mu\nu}^\cA V_\cA \c^\adot \c^{\mu\nu}\,\epsilon\ ,
\nn\w2
\delta\chi^a &=& \ft14 P_\mu^{a\adot} \c^\adot \c^\mu \epsilon +\ft1{12} H_{\mu\nu\rho}^a \c^{\mu\nu\rho}\epsilon +\ft14 F_{\mu\nu}^\cA {\tilde V}_A \c^a\c^{\mu\nu}\,\epsilon\ ,
\nn\w2
\delta A_\mu^\cA &=& -{\bar\epsilon} V^\cA \psi_\mu + {\bar\psi}_\mu V^\cA \epsilon
+\ft12 {\bar\epsilon}\c_\mu\c^a V^\cA \chi^a +\ft12 {\bar\chi}^\adot V^\cA \c^\adot \c_\mu\epsilon\ ,
\nn\w2
\Delta B_{\mu\nu m} &=& \s2\,\cV_m^a \left( {\bar\psi}_{[\mu} \c_{\nu]} \c^a\epsilon +\ft12{\bar\chi}^a\c_{\mu\nu}\epsilon\right) - \s2\,\cV_m^\adot \left( {\bar\psi}_{[\mu} \c_{\nu]} \c^\adot\epsilon +\ft12{\bar\chi}^\adot\C_{\mu\nu}\epsilon\right)\ ,
\nn\w2
\delta V_\cA &=& \ft12 \left(\c_a V_\cA\c_\adot\right) \left({\bar\chi}_a\c_\adot\epsilon
+{\bar\chi}_\adot\c_a\epsilon\right)\ ,
\label{susy1}
\eea
where $\Delta B_{\mu\nu}$ is the gauge covariant variation defined in Appendix B,
\be
D_\mu\epsilon= \partial_\mu \epsilon +\ft14 \omega_\mu{}^{rs}\c_{rs}\epsilon + \ft14 Q_\mu{}^{ab}\c_{ab}\epsilon\ ,
\ee
and $``\sim"$ denotes transposition. The chiralities are shown explicitly only when there is an ambiguity. Otherwise, when suppressed, they can easily be deduced from the structure of the terms (see Appendix A for notation and conventions).

\section{Gauging $G_0 \subset  {\rm SO}(5,5)$ }\label{sec:gauged}

Using the embedding tensor formalism \cite{Nicolai:2000sc,deWit:2002vt,deWit:2005hv}, we will find the most general gauging of a group $G_0 \subset {\rm SO}(5,5)$ by employing a suitable combination of the $16$ vector fields in the  theory.

\subsection{The Embedding Tensor}

The key ingredient in the construction is the covariant derivative
\bea
D_{\mu}&=&
\partial_{\mu}-gA_\mu{}^\cA\,\Theta_\cA{}^{MN}\,t_{MN}\ , \label{et}
\eea
with ${\rm SO}(5,5)$ generators $t_{MN}=t_{[MN]}$ and an embedding tensor $\Theta_\cA{}^{MN}$.
It follows from supersymmetry \cite{deWit:2002vt} that the latter can be parametrized in terms of a tensor $\theta^{B M}$ transforming in the ${\bf 144}_{c}$ representation, i.e.\ satisfying
\be
\c_{M\cA\cB}\,\theta^{\cB M}=0\ ,
\ee
as follows:
\be \Theta_\cA{}^{MN}=-\theta^{\cB[M}\,\c^{N]}{}_{\cB\cA} \equiv \left(\c^{[M} \theta^{N]}\right)_\cA\ . \ee
In this paper, we show that indeed every embedding tensor in the ${\bf 144}_c$ (which also satisfies the quadratic constraints \eq{quad} below) defines a  consistent gauging, and present the full Lagrangian.

The ${\rm SO}(5,5)$ algebra is realized by generators $t_{{MN},K}{}^{L}= 4\eta_{K[M}\delta_{N]}^{L}$\,
in the vector representation and $t_{{MN},A}{}^B=(\gamma_{{MN}})_A{}^B$
on the spinor representation, respectively, satisfying
\bea
{}[\,t_{KL},t_{MN}]&=& 4\,(\eta_{K[M}\,t_{N]L}-\eta_{L[M}\,t_{N]K})\;.
\eea
Therefore, the gauge algebra generators $X_\cA= \theta_\cA{}^{MN} t_{MN}$ take the form
\bea
X_{\cA\cB}{}^\cC &\!=\!&  (\c^M\theta^N)_\cA (\c_{MN})_\cB{}^\cC\ ,
\quad
X_{\cA,M}{}^{{N}}= 2 (\c_M\theta^N)_\cA +2 (\c^N\theta_M)_\cA\ ,
\eea
acting on spinors and vectors, respectively. The quadratic constraints on the embedding tensor state that
\bea
{}[X_\cA,X_\cB\,] &=& -X_{\cA\cB}{}^\cC \,X_\cC\ .
\eea
Some computation shows that this reduces to imposing
\bea
\theta^{\cA M}\,\theta^{\cB N}\,\eta_{MN}&=& 0\;,
\qquad
\theta^{\cA M}\,\theta^{\cB [N}\,
(\gamma^{P]})_{\cA\cB}~=~0\;,
\label{quad}
\eea
on the tensor $\theta^{{\cal A}M}$. This means that the quadratic constraints
transform in the ${\bf 10}+{\bf 126}_{c}+{\bf 320}$ of $SO(5,5)$ ---
and thus in the representation conjugate to the 6-forms of the theory~(\ref{hpf}).
They ensure, for example, that
\bea
\theta^{\cA M}\,X_{\cA\cB}{}^\cC &=& 0\;.
\eea
The generators $X_{\cA\cB}{}^\cC$ satisfy
\bea
X_{(\cA\cB)}{}^{\cC}&=&
-\theta^{D M}\,(\c^{N})_{D(\cA}\,
(\c_{MN})_{\cB)}{}^\cC
~=~ -(\c_{M})_{\cA\cB}\,\theta^{\cC M}
~\equiv~
d_{M,\cA\cB} \,Z^{\cC,M}
\;,
\label{Xsym}
\eea
where we have introduced the general notation
\be
d_{M,\cA\cB}\equiv(\c_M)_{\cA\cB}\ ,\qquad  Z^{\cC,M}\equiv-\theta^{\cC \cM}\ .
\ee
As we have discussed in the introduction, gauging the theory in general not only corresponds to covariantizing the derivatives according to (\ref{et}) but also induces a nontrivial deformation of the hierarchy of $p$-form tensor gauge transformations. In particular, $p$-forms start to transform by (St\"uckelberg)-shift under the gauge transformations of the $(p\!+\!1)$-forms. The corresponding tensors required to intertwine between the representations of $p$- and $(p\!+\!1)$-forms are generated by the embedding tensor. The lowest intertwining tensors can be obtained by evaluating the general formulas of~\cite{deWit:2005hv} for our case, yielding
\bea
Z^{\cC,M}&\equiv&-\theta^{\cC \cM}
\nonumber\\[1ex]
Y_{M,\cA}{}^{N} &=&
-2\,\theta^{\cB P}\,(\c^{Q})_{\cB\cA}
\, \eta_{MP}\,\delta_{Q}^{N} + 2\,\theta^{\cB P}\,(\c^{Q})_{\cB\cA}\,\eta_{MQ}\,
\delta_{P}^{N}-2\, \theta^{\cB N}\, (\c_{M})_{\cB\cA}\nonumber\\
&=& -2\,\theta^{\cB K} \,\eta_{MK}\, (\c^{N})_{\cA\cB} \ ,
\label{YZ}
\eea
for the tensors intertwining between vectors/2-forms and 2-/3-forms, respectively. In particular, the latter tensor encodes the representation content of 3-forms, required for consistency of the deformed tensor gauge algebra. As the 3-forms (with the generic index structure $C_{\mu\nu\rho\,N}{}^\cA$) will always appear under projection $Y_{M,\cA}{}^{N}\,C_{\mu\nu\rho\,N}{}^\cA$, the particular form of (\ref{YZ}) shows that out of this general set only the 16 projected 3-forms $(\c^{N})_{\cA\cB} \,C_{\mu\nu\rho\,N}{}^\cA\equiv C_{\mu\nu\rho\,\cB}\,$ enter the theory. This is in accordance with the field content discussed in the introduction, in particular with the fact that as a consequence of their on-shell duality, 3-forms should transform in the representation conjugate to the vector fields. With (\ref{YZ}), the $p$-form tensor gauge algebra in six dimensions can now be written down by evaluating the general formulas of~\cite{deWit:2005hv} (see in particular \cite{Samtleben:2005bp}, Appendix~A).

General $p$-form variations are most conveniently expressed in terms of the ``covariant variations''
\footnote{Note that $B$ and $\Xi$ have been rescaled by a factor of $\s2$, and $\theta$ by a minus sign, w.r.t. the formulae provided in \cite{Samtleben:2005bp}.}
\bea
\Delta A_\mu^A
&\equiv& \delta A_\mu{}^A
\;,\\[1ex]
\Delta B_{\mu\nu\,M} &\equiv&
\delta B_{\mu\nu\,M}
- \s2\,(\gamma_{M})_{AB}\,A_{[\mu}{}^A\,\delta A_{\nu]}{}^B
\;,
\nonumber\\[1ex]
\Delta C_{\mu\nu\rho\,A}&\equiv&
\delta C_{\mu\nu\rho\,A}
-3\s2\,(\gamma^{M})_{AB}\,B_{[\mu\nu\,M} \,\delta A_{\rho]}{}^B
- 2  (\gamma^{M})_{AB}(\gamma_{M})_{CD}\,
A_{[\mu}{}^B A_{\nu}{}^C\,\delta A_{\rho]}{}^D
\;.
\nonumber
\label{D1}
\eea
The full non-abelian gauge transformations are then given by\footnote{
As usual in even dimensions there is a subtlety with the gauge transformation law
of the $D/2$-forms~\cite{deWit:2005ub,Samtleben:2007an}
requiring that eventually in the off-shell formulation of gauge transformations,
${\cal H}_{\mu\nu\rho\,M}$ in the last line of (\ref{gauge}) is replaced by
${\cal G}_{\mu\nu\rho\,M}$ from (\ref{d3}), below. }
\bea
\Delta A_\mu{}^A &=&
D_\mu\Lambda^A
+ \s2\, g\,\theta^{AM}\,\Xi_{\mu\,M}
\;,
\nonumber\\[1ex]
\Delta B_{\mu\nu\,M}
&=&
2\,D_{[\mu} \Xi_{\nu]M}
-\s2\,(\gamma_{M})_{AB}\, \Lambda^A\,{\cal H}_{\mu\nu}^B
+  \s2\,g\,\theta^{AN} \,\eta_{MN}\,\Phi_{\mu\nu\,A}
\;,
\nonumber\\[1ex]
\Delta C_{\mu\nu\rho\,A}
&=&
3\,D_{[\mu}\Phi_{\nu\rho]\,A}
 + 3\s2\, (\gamma^{M})_{AB}\,{\cal H}_{[\mu\nu}^B\,\Xi_{\rho]M}
+ \s2\,(\gamma^{M})_{AB}\,\Lambda^B\,{\cal H}_{\mu\nu\rho\,M}
\;,
\label{gauge}
\eea
with gauge parameters $\Lambda^A$, $\Xi_{\mu M}$, $\Phi_{\mu\nu\,A}$,
and the covariant field strengths
\bea
{\cal H}_{\mu\nu}^A &\equiv&
2\,\partial_{[\mu} A_{\nu]}{}^A
+g\,X_{[BC]}{}^A \,A_\mu{}^B A_\nu{}^C- \s2\, g\,\theta^{AM} \,B_{\mu\nu\,M}
\;,
\nonumber\\[2ex]
{\cal H}_{\mu\nu\rho\,M} &\equiv&
3\, D_{[\mu} B_{\nu\rho]\,M} +
3\s2\,(\gamma_{M})_{AB}\,A_{[\mu}{}^A\,\Big(\partial_{\nu} A_{\rho]}{}^B+
\ft13 g X_{[CD]}{}^B A_{\nu}{}^C A_{\rho]}{}^D\Big)
\nn\\
&&{}
- \s2\,g\,\theta^{AN}\,\eta_{MN}\, C_{\mu\nu\rho\,A} \;,
\nonumber\\[2ex]
{\cal H}_{\mu\nu\rho\lambda\,A} &\equiv&  4D_{[\mu} C_{\nu\rho\lambda]\,A} -(\c^M)_{AB} \Big( 6\s2\, B_{\mu\nu M} \cH_{\rho\lambda}^B+6 g\theta^{BN} B_{[\mu\nu M} B_{\rho\lambda] N}
\nn\w2
&& +8(\c_M)_{CD} A_{[\mu}^B A_\nu^C \partial_\rho A_{\lambda]}^D +2 (\c_M)_{CF} X_{DE}{}^F A_{[\mu}{}^B A_\nu{}^C A_\rho{}^D A_{\lambda]}{}^E\Big)\ .
\label{covAH}
\eea
Under arbitrary variations these field strengths transform as
\bea
\delta\,{\cal H}_{\mu\nu}^A &=&
2\,D_{[\mu}\, (\Delta A_{\nu]}{}^A)
 -\s2 g\, \theta^{AM} \,\Delta B_{\mu\nu\,M}
 \;,\nn\\[3ex]
 \delta\,{\cal H}_{\mu\nu\rho\,M} &=&
 3\, D_{[\mu} (\Delta B_{\nu\rho]\,M})
+ 3\s2\,(\gamma_{M})_{AB}\,{\cal H}_{[\mu\nu}^A \, \Delta A_{\rho]}{}^B
-\s2\,g\,\theta^{AN}\,\eta_{MN}\,
\Delta C_{\mu\nu\rho\,A} \;.\nn
\\[3ex]
\delta \cH_{\mu\nu\rho\lambda\,A} &=& 4D_{[\mu} \Delta C_{\nu\rho\lambda] A} -4(\c^M)_{AB} \left(\ft3{\s2}\,\cH_{[\mu\nu}^B \Delta B_{\rho\lambda] M} -\s2\, \cH_{[\mu\nu\rho\,M} \Delta A_{\lambda]}{}^B\right)\ .
\eea
One of the consequences of the gauge covariantization \`a la \eq{et} is the modification of the scalar currents as
\be
P_\mu^{a\adot}\ \ \ \rightarrow \ \ \ \cP_\mu^{a\adot} = \ft14 {\bar V} \c^a\c^\adot \cD_\mu V\ ,
\qquad
Q_\mu^{ab} \ \ \ \rightarrow \ \ \ \cQ_\mu^{ab} = \ft18 {\bar V} \c^{ab} \cD_\mu V\ ,
\ee
and similarly for $Q_\mu{}^{\adot\bdot}$, with the gauge covariant derivative given by
\be
\cD_\mu V = D_\mu V - g\,({\bar A}_\mu \c^M\theta^N)\,\c_{MN} V\ .
\ee
This leads to the following modified integrability equations:
\bea
&& \cD_{[\mu} \cP_{\nu]}^{a\adot} +4g \cF_{\mu\nu}^{\a\ad}\,T_{\a\ad}^{a\adot}=0\ ,
\nn\w2
&&
\cQ_{\mu\nu}^{ab}+ \ft12 \cP_{[\mu}^{a\adot}\cP_{\nu]}{}^{\adot b} +4g \cF_{\mu\nu}^{\a\ad}\,T_{\a\ad}^{ab}=0\ ,
\nn\w2
&&
\cQ_{\mu\nu}^{\adot\bdot}+ \ft12 \cP_{[\mu}^{a\adot}\cP_{\nu]}{}^{a\bdot } +4g \cF_{\mu\nu}^{\a\ad}\,T_{\a\ad}^{\adot\bdot}=0\ ,
\label{4}
\eea
where $\cQ_{\mu\nu}^{ab} = 2\partial_{[\mu}\cQ_{\nu]}^{ab}+ 2\cQ_{[\mu}{}^{ac} \cQ_{\nu]c}{}^b$ and
\bea
\cF_{\mu\nu}{}^\cA &\equiv&
2\,\partial_{[\mu} A_{\nu]}{}^\cA
+ g\,X_{[\cB\cC]}{}^\cA \,A_\mu{}^\cB  A_\nu{}^\cC
\nn\w2
&=& 2\,\partial_{[\mu} A_{\nu]}{}^\cA + g (\c_{MN} A_{[\mu})^A {\bar\theta}^M \c^N A_{\nu]}\ ,\qquad
\eea
and
\bea
T^{a\adot}_{\a\ad} &=& \ft1{32} \left({\bar V}_{\a\ad}\c_M\theta_N\right)\,\left({\bar V}\c^a\c^\adot \c^{MN} V\right)\ ,
\nn\w2
T^{ab}_{\a\ad} &=& \ft1{32} \left({\bar V}_{\a\ad}\c_M\theta_N\right)\,\left({\bar V}\c^{ab} \c^{MN} V\right)\ ,
\nn\w2
T^{\adot\bdot}_{\a\ad} &=& \ft1{32} \left({\bar V}_{\a\ad}\c_M\theta_N\right)\,\left({\bar V}\c^{\adot\bdot} \c^{MN} V\right)\ .
\label{t1}
\eea
These expressions can be simplified and their group theoretical meaning can be made more transparent by making use of \eq{gr2} and recalling that $\c_M\theta^M=0$. As a result, we find
\bea
&& T^{ab} = \c^{[a}\,T^{b]}\ ,\qquad   T^{\adot\bdot}=- T^{[\adot} \c^{\bdot]}\ ,\qquad T^{a\adot}= -\ft12 (\c^a T^\adot +T^a \c^\adot)\ ,
\nn\w2
&&\c^a T^a +T^\adot \c^\adot=0\ ,
\label{t2}
\eea
where we have defined the {\it T-tensors},
\be
T^a= \cV_M{}^a\,\theta^{AM} V_A\ ,\qquad T^\adot = -\cV_M{}^\adot \,\theta^{AM} V_A\ .  \label{t3}
\ee
Thus, $T^\unA =(T^a,T^\adot)$ is in one-to-one correspondence with the embedding tensor $\theta^M$. For later purposes, it is convenient to also define
\be
T\equiv \c^a T^a =  -T^\adot \c^\adot\ .
\ee
The quadratic constraints \eq{quad} translate into
\be
T^a_{\a\ad} T^a_{\b\bd} - T^\adot_{\a\ad} T^\adot_{\b\bd}=0\ ,\qquad  T^{\unC\, \a\ad}\,\c^{[\unA}_{\a\ad,\b\bd}\,T^{\unB]\b\bd} =0\ ,
\label{qc2}
\ee
where $\c^\unA= (\c^a \times 1, 1\times \c^\adot)$. Restricting to ${\rm SO}(5)_I\times {\rm SO}(5)_{II}$ directions, several identities result from the latter equation. For example, restriction to the ${\rm SO}(5)_I$ direction, upon the use of \eq{cc5}, gives
\be
T^{(a} {\tilde T}^{b)} - \ft14 \tr (T^a {\tilde T}^b) = \ft14 \c_c\,\tr (T^a {\tilde T}^c \c^b) \ .
\label{TTe}
\ee
We recall that  $``\sim"$ denotes transposition. The nontrivial content of this equation is the antisymmetric
part in its free $SO(5)$ indices, namely,
\be
\tr ({\tilde T}^c \c^{[a} T^{b]})=0\ ,
\label{master}
\ee
while the symmetric projection, contains no new information, in view of \eq{cc6}. A useful identity needed in establishing the supersymmetry of the Lagrangian is obtained by evaluating the
antisymmetric part of $\c^a T{\tilde T}^a$. Using the trace of the constraint equation \eq{TTe}, and recalling \eq{cc6}, we obtain
\be
\c^a T{\tilde T}^a +T^a \tilde T \c^a = 4 T^a{\tilde T}^a - \tr\, T^a{\tilde T}^a -2 T{\tilde T}
+\tr\,T{\tilde T} \ .
\ee
Next, we observe that the constraint \eq{qc2} enables us to covariantize the identities \eq{4}
\bea
&& \cD_{[\mu} \cP_{\nu]}^{a\adot} -4g \cH_{\mu\nu}^A\,\tr\, T^{a\adot} {\tilde V}_A =0\ ,
\label{4b}\w2
&& \cQ_{\mu\nu}^{ab} + \ft12 P_{[\mu}^{a\adot}P_{\nu]\adot}{}^b -4g \cH_{\mu\nu}^A\,\tr\, T^{ab} {\tilde V}_A  =0\ .
\label{4c}
\eea
Further useful relations are furnished by the derivatives of the $T$-tensors, which take the form
\bea
\cD_\mu T^a &=& \ft14\, \cP_\mu^{b\bdot} \left(\c^b T^a\c^\bdot -2\delta^{ab} T^\bdot \right)\ ,
\nn\w2
\cD_\mu T^\adot &=& \ft14\, \cP_\mu^{b\bdot} \left(\c^b T^\adot\c^\bdot -2\delta^{\adot\bdot} T^b \right)\ ,
\nn\w2
\cD_\mu T &=& \ft12\, \cP_\mu^{a\adot} \left( -\c^a T^\adot + T^a\c^\adot -\ft12 \c^a T\c^\adot\right)\ .
\eea
The quantities $\cP$ and $\cQ$ can conveniently be written as
\bea
\cP_\mu^{a\adot} &=& P_\mu^{a\adot} +8g\, A_\mu^A~\tr\, T^{a\adot} {\tilde V}_A\ ,
\label{pqg1}\w2
\cQ_\mu^{ab} &=& Q_\mu^{ab} +4g\, A_\mu^A~\tr\, T^{ab} {\tilde V}_A\ ,
\label{pqg2}
\eea
and similarly for $\cQ_{\mu\adot\bdot}$. Finally, the modified Bianchi identities are
\bea
\cD_{[\mu}\cH_{\nu\rho]}^A&=& -\ft{\s2}{3}\,g\,\theta^{AM} \cH_{\mu\nu\rho M} \ ,
\label{B1}\w2
\cD_{[\mu} \cH_{\nu\rho\sigma] M} &=& \ft3{2\s2}\, {\bar\cH}_{[\mu\nu} \c_M \cH_{\rho\sigma]}
-\ft1{2\s2}\,g\, \theta_M^A\cH_{\mu\nu\rho\sigma\,A}\ .
\label{B2}
\eea

\subsection{The Gauged Maximal D$=$6 Supergravity}

The building blocks we have just described can now be used to gauge the maximal D$=$6 supergravity. Thus, we introduce the magnetic potentials $B_{\mu\nu}^m$ and the 3-form potentials $C_{\mu\nu\rho A}$ accordingly, and in the ungauged Lagrangian we make the replacements
\be
H_{\mu\nu\rho m} \rightarrow \cH_{\mu\nu\rho m}\ ,\qquad H_{\mu\nu}^A \rightarrow \cH_{\mu\nu}^A\ ,\qquad P_\mu^{a\adot} \rightarrow \cP_\mu^{a\adot}\ ,
\ee
as well as gauge covariantize the derivatives by the prescription
\be
D_\mu \rightarrow \cD_\mu\ , \qquad Q_\mu^{ab} \rightarrow \cQ_\mu^{ab}\ ,\qquad  Q_\mu^{\adot\bdot} \rightarrow \cQ_\mu^{\adot\bdot}\ ,
\ee
in the supersymmetry transformation rule, and the Lagrangian with the exception of the topological terms. They are modified by the requirement of all the gauge symmetries described in the previous section. This turns out to be highly constraining nontrivial requirement which  remarkably fixes the topological terms entirely, as will be described in detail in the next section.
These modifications will introduce new, gauge coupling constant $g$-dependent supersymmetry variations due to the explicitly  $g$-dependent terms in \eq{4b}, \eq{pqg1}, \eq{pqg2} and \eq{B2}. To cancel them, as usual, we parametrize the most general fermionic mass terms that are linear in the T-tensors, and a potential that is quadratic in the T-tensors, and introduce linear in the T-tensor terms in the supersymmetry variations of the fermions. As for the supersymmetry transformations of the newly introduced higher rank $p$-forms, that of $B_{\mu\nu}^m$ is straightforward by simply requiring that together with $B_{\mu\nu m}$ they form a 10-plet of ${\rm SO}(5,5)$. Regarding the 3-form potential $C_{\mu\nu\rho A}$ we simply parameterize its supersymmetry transformation rules in a fashion dictated by gauge symmetries and dimensional analysis. Requiring that all the $g$-dependent variations cancel, we determine all the coefficients used in parameterizing the Lagrangian and supersymmetry transformation rules. The subtle features that arise in these computations are to a large extent parallel to those encountered in the construction of the gauged maximal supergravities in D$=$4 \cite{de Wit:2007mt}. We will spell out some more details of the salient features in this computation but first, let us present our results.

We have found that the Lagrangian $\cL=\cL_B+\cL_F$, up to quartic fermion terms, is given by
\bea
e^{-1} \cL_B &=&  \ft14 R  -\ft1{12} \cH_m \cdot K^{mn} \cH_n
-\ft14 \cM_{AB} \cH_{\mu\nu}^\cA \cH^{\mu\nu\cB}
\nn\w2
&& -\ft1{16} \cP_\mu^{a\adot} \cP^\mu_{a\adot}  + g^2 \left(
\tr\,T^a{\tilde T}^a -\ft12\,\tr\,T{\tilde T} \right) +e^{-1} \cL_{top}\ ,
\eea
where $\cL_{\rm top}$ is the topological part of the Lagrangian given in the next section, and
\bea
e^{-1} \cL_F &=&
 -\ft12 {\bar\psi}_{+\mu} \c^{\mu\nu\rho} \cD_\nu \psi_{+\rho}
-\ft12 {\bar\psi}_{-\mu} \c^{\mu\nu\rho} \cD_\nu \psi_{-\rho}
-\ft12 {\bar\chi}^a \c^\mu \cD_\mu \chi^a
-\ft12 {\bar\chi}^\adot\c^\mu \cD_\mu \chi^\adot \nn\w2
&&
+\ft1{4}\cP_\mu^{a\adot}\,I^\mu_{a\adot} - \ft12 \cH_{\mu\nu}^\cA\,J^{\mu\nu}_\cA
+\ft16 \left(P_+ \cH^a \cdot \cO^a + P_-\cH^\adot \cdot \cO^\adot \right)
\nn\w2
&& + g{\bar\psi}_{+\mu}\c^{\mu\nu} T\psi_{-\nu} + 2g\left({\bar\psi}_\mu\c^\mu T^a\chi^a +{\bar\chi}^\adot T^\adot \c^\mu\psi_\mu\right)
\nn\w2
&&  +\ft12 g\left({\bar\chi}^\adot T\c^\adot\c^\mu\psi_\mu - {\bar\psi}_\mu\c^\mu \c^a T \chi^a\right)
\nn\w2
&& + g{\bar\chi}^\adot \left( 2\c^a T^\adot - 2T^a\c^\adot + \ft12\,\c^a T\c^\adot \right) \chi^a\ ,
\eea
where the fermionic bilinears are as given in \eq{fb}.

The supersymmetry transformations are
\bea
\delta e_\mu{}^r &=& {\bar\epsilon}_+\c^r \psi_{+\mu}+ {\bar\epsilon}_-\c^r\psi_{-\mu}\ ,
\nn\w2
\delta\psi_{\mu +} &=& \cD_\mu \epsilon_+ -\ft1{24}\cH_{\rho\sigma\kappa}^a\c^a\c^{\rho\sigma\kappa}\c_\mu\epsilon_+
+\ft18 \left(\c_\mu{}^{\nu\rho}-6\d_\mu^\nu\c^\rho\right) \cH_{\nu\rho}^\cA V_A\, \epsilon_-
+ \ft14 g \,\c_\mu T\epsilon_-\ ,
\nn\w2
\delta\psi_{\mu -}&=& \cD_\mu \epsilon_- -\ft1{24}\cH_{\rho\sigma\kappa}^\adot\c^\adot\c^{\rho\sigma\kappa}\c_\mu\epsilon_-
+\ft18 \left(\c_\mu{}^{\nu\rho}-6\d_\mu^\nu\c^\rho\right) \cH_{\nu\rho}^\cA {\tilde V}_A\, \epsilon_+
-\ft14 g\,\c_\mu {\tilde T}\epsilon_+\ ,
\nn\w2
\delta\chi^\adot &=& \ft14 \cP_\mu^{a\adot} \c^a \c^\mu \epsilon +\ft1{12} \cH_{\mu\nu\rho}^\adot \c^{\mu\nu\rho}\epsilon +\ft14 \cH_{\mu\nu}^\cA V_A \c^\adot\c^{\mu\nu}\,\epsilon + 2g\,T^\adot \epsilon
+ \ft12 g \, T\c^\adot \epsilon\ ,
\nn\w2
\delta\chi^a &=& \ft14 \cP_\mu^{a\adot} \c^\adot \c^\mu \epsilon +\ft1{12} \cH_{\mu\nu\rho}^a \c^{\mu\nu\rho}\epsilon +\ft14 \cH_{\mu\nu}^\cA {\tilde V}_A \c^a\c^{\mu\nu}\,\epsilon + 2g \,{\tilde T}^a \epsilon - \ft12 g\,{\tilde T}\c^a \epsilon\ ,
\nn\w2
\delta A_\mu^\cA &=& -{\bar\epsilon} V^\cA \psi_\mu + {\bar\psi}_\mu V^\cA \epsilon
+\ft12 {\bar\epsilon}\c_\mu\c^a V^\cA \chi^a +\ft12 {\bar\chi}^\adot V^\cA \c^\adot \c_\mu\epsilon\ ,
\nn\w2
\Delta B_{\mu\nu M} &=& \s2\,\cV_M^a \left( {\bar\psi}_{[\mu} \c_{\nu]} \c^a\epsilon +\ft12{\bar\chi}^a\c_{\mu\nu}\epsilon\right) -\s2\,\cV_M^\adot \left( {\bar\psi}_{[\mu} \c_{\nu]} \c^\adot\epsilon +\ft12{\bar\chi}^\adot\c_{\mu\nu}\epsilon\right) \ ,
\nn\w2
\Delta C_{\mu\nu\rho A} & =& 3\left( {\bar\epsilon}V_A \c_{[\mu\nu} \psi_{\rho]}-{\bar\psi}_{[\mu} \c_{\nu\rho]} V_A\epsilon \right)  + \ft12 \left({\bar\epsilon} \c^a V_A \c_{\mu\nu\rho} \chi^a + {\bar\chi}^\adot V_A \c^\adot \c_{\mu\nu\rho}\epsilon\right)\ ,
\nn\w2
\delta V_\cA &=& \ft12 \left(\c^a V_\cA\c^\adot\right) \left({\bar\chi}^a\c^\adot\epsilon
+{\bar\chi}^\adot\c^a\epsilon\right)\ .
\label{susy2}
\eea
We emphasize again that the $\pm$ chiralities have been shown explicitly only when necessary, and when suppressed they can be deduced from the structure of the terms. We also note that $\cH_{\mu\nu\rho a}$ and $\cH_{\mu\nu\rho\adot}$ are defined by
\be
\cH_m = \s2\,\left(\cV_m{}^a P_+ \cH^a - \cV_m{}^\adot P_- \cH^\adot\right)\ ,
\ee
where we have suppressed the tensorial indices. This is analogous to the relation \eq{hm1} in the ungauged model. Similarly, we can define the analog of the field strength \eq{d31} as
\be
\cG_M = \left(\begin{array}{c}  \cG_m\\   \cG^m \\ \end{array}\right) =
\left(\begin{array}{c}  \cH_m\\  j K^{mn} \cH_n \\ \end{array}\right)\ .
\label{d3}
\ee
As in \eq{ha1}, it follows that

\be
  P_+ \cH_a = \ft1{\s2}\,P_+ \cG_M \cV^M_a\ , \qquad P_- \cH_\adot = -\ft1{\s2}\,P_-\cG_M \cV^M_\adot\ .
\ee
Thus, the supersymmetry transformations, as well as the Pauli couplings involving $\cG$, are manifestly duality-covariant. The supersymmetry algebra is expected to close on-shell with field dependent composition symmetry parameters, as usual.  Normally, the fermionic field equations are needed for the closure, but here, the closure on the three-form potential requires its field equation as well. In the next section, we will show that this field equation takes the simple form $\theta^{AM}\,(\cG_M-\cH_M)=0$.

We conclude this section by expressing the potential explicitly in terms of the embedding tensor and the coset representatives, and observe that it takes the remarkably simple form
\be
V(\phi)= \ft12~\theta^{AM}\theta^{BN} \cV_M^a\cV_N^b\left({\bar V}_A \c^b\c^a V_B\right)\ .
\ee

\subsection{The Topological Term}

In establishing the gauge and supersymmetry of the action a highly complicated topological term is needed.
The full topological term is given by
\bea
\cL_{\rm top} &=& \frac{1}{36} e^{-1} \epsilon^{\mu\nu\rho\sigma\kappa\lambda} \Big[
-\tb_m C_{\mu\nu\rho} \left(\tb^m C_{\sigma\kappa\lambda}+\s2\,\cH_{\sigma\kappa\lambda}{}^m\right)
-\frac{3}{\s2}\,\tb^M\c^N\theta^P B_{\mu\nu M} B_{\rho\sigma N} B_{\kappa\lambda P}
\nn\w2
&&
+18 \left( \tb_M\c^mA_\mu  \partial_\nu B_{\rho\sigma m} + \tb_m\c_M A_\mu \partial_\nu B_{\rho\sigma}{}^m\right) B_{\kappa\lambda}{}^M
+18 \ab_\mu\c_m\theta^M \ab_\nu \c^m \theta^N B_{\rho\sigma M} B_{\kappa\lambda N}
\nn\w2
&&
18\ab_\mu\c^M\theta_m \left(\ab_\nu\c^N\theta^m -2\ab_\nu\c^m\theta^N\right) B_{\rho\sigma M} B_{\sigma\kappa N}
-9\s2 \partial_\mu\ab_\nu \c^m \partial_\rho A_\sigma B_{\kappa\lambda m}
\nn\w2
&&
-18\s2 \left(\tb^M\c_m A_\mu - \tb_m\c^M A_\mu\right) \partial_\nu \ab_\rho\c^m A_\sigma\,B_{\kappa\lambda M}
-3\s2\,\ab_\mu\c^m X_{\nu\rho} \partial_\sigma B_{\kappa\lambda m}
\nn\w2
&&
+6\s2\,\tb^M\c_N A_\kappa\,\partial_\rho\ab_\sigma \c^N A_\lambda\,B_{\mu\nu M}
-\frac{3}{\s2}\,\ab_\mu\c^N X_{\rho\sigma}\,\ab_\nu\c_N\theta^M\,B_{\kappa\lambda M}
\nn\w2
&&
-6\s2\,\ab_\mu\c^m X_{\nu\rho} \left(\ab_\sigma\c^M\theta_m-\ab_\sigma\c_m\theta^M\right) B_{\kappa\lambda M}
+9\ab_\mu\c_m\partial_\nu A_\rho \ab_\sigma\c^m \partial_\kappa A_\lambda
\nn\w2
&& +\frac{12}{5}\ab_\mu\c_M X_{\nu\rho}\, \ab_\sigma \c^M\partial_\kappa A_\lambda
-6\ab_\mu\c^m X_{\nu\rho} \left( \ab_\sigma \c_m\partial_\kappa A_\lambda
-\frac16\ab_\sigma\c_m X_{\kappa\lambda}\right)\Big]\ ,
\label{Ltop}
\eea
where the gauge coupling constant $g$ is suppressed and
\be
X_{\mu\nu}{}^A \equiv \ab_{[\mu}\c^M\theta^N (\ab_{\nu]}\c_{MN})^A\ .
\ee
The topological Lagrangian ${\cL}_{top}$ is completely fixed by requiring gauge invariance of ${\cal L}_{\rm top}+{\cal L}_{\rm kin}$. In fact, the topological term can already completely be determined just starting from its leading term $\theta_m^{A}\,C_{\mu\nu\rho\,{A}}\,\partial_{\sigma}B_{\kappa\lambda}{}^{m}$ and completing the term by requiring invariance under tensor gauge transformations $\delta_\Phi {\cal L}_{\rm top}=0=\delta_\Xi {\cal L}_{\rm top}$. Subsequently, one can show that its general variation takes the fully covariant form \eq{varLtop} below, which is a strong consistency check. Useful identities needed for these computations are provided in Appendix B.

Note that the mass term for the three forms $\theta_m^A\,\theta^{Bm}$ is automatically antisymmetric due to the quadratic constraint. Moreover, no such term would exist with full ${\rm SO}(5,5)$ covariance, i.e.\ it is essential here that the Lagrangian exhibits only ${\rm GL}(5)$ covariance. Also the cubic $B^{3}$ coupling ${\bar\theta}^M \c^N \theta^P$ is automatically symmetric in $(MNP)$ due to the quadratic constraint \eq{quad}. Finally, note that also the $A^{6}$ term could not exist in an ${\rm SO}(5,5)$ covariant Lagrangian: there is no ${\rm SO}(5,5)$ singlet in the tensor product of $\Theta^{2}A^{6}$. Again it is essential that ${\rm SO}(5,5)$ is broken to ${\rm GL}(5)$.

In the ungauged theory ($\theta^{AM}=0$) the topological term \eq{Ltop} is simply
\be
\cL = \ft12 \bA  \c^m dA \wedge \left( dB_m +\ft14\bA  \c_m dA\right) \ ,
\label{Ltopungauged}
\ee
and contained in \eq{LB}. For electric gaugings ($\theta_{m}^{A}=0$) the topological term reduces to:
\bea
\cL_{\rm top} &=& \frac{1}{36} e^{-1}\epsilon^{\mu\nu\rho\sigma\kappa\lambda} \Big[
-\frac3{\s2}\,\tb^m\c^n\theta^p B_{\mu\nu m} B_{\rho\sigma n} B_{\kappa\lambda p}
+18\tb^n\c^m A_\mu  \partial_\nu B_{\rho\sigma m} B_{\kappa\lambda n}
\nn\w2
&&
+18\ab_\mu\c_m\theta^n \ab_\nu \c^m \theta^p B_{\rho\sigma n} B_{\kappa\lambda p}
-9\s2\, \partial_\mu\ab_\nu \c^m \partial_\rho A_\sigma B_{\kappa\lambda m}
\nn\w2
&&
-18\s2\, \tb^n\c_m A_\mu \partial_\nu \ab_\rho \c^m A_\sigma\,B_{\kappa\lambda n}
-3\s2\,\ab_\mu\c^m X_{\nu\rho} \partial_\sigma B_{\kappa\lambda m}
\nn\w2
&&
+6\s2\,\tb^m\c_N A_\mu \left(\ab_\nu\c^N\partial_\rho A_\sigma -\frac14 \ab_\nu \c^N X_{\rho\sigma}\right) B_{\kappa\lambda m}
\nn\w2
&&
+6\s2\,\ab_\mu\c^m X_{\nu\rho} \ab_\sigma\c_m\theta^n B_{\kappa\lambda n} +9\ab_\mu\c_m\partial_\nu A_\rho \ab_\sigma\c^m \partial_\kappa A_\lambda \nn\w2
&& +\frac{12}{5}\ab_\mu\c_M X_{\nu\rho}\, \ab_\sigma \c^M\partial_\kappa A_\lambda
-6\ab_\mu\c^m X_{\nu\rho} \left( \ab_\sigma \c_m\partial_\kappa A_\lambda
-\frac16\ab_\sigma\c_m X_{\kappa\lambda}\right)\Big]
\label{Ltopelectric}
\eea
and one sees explicitly that in this case neither 3-forms $C_{{\mu\nu\rho\,A}}$ nor magnetic two-forms $B_{\mu\nu}{}^{m}$ enter this Lagrangian.
For the $B^{3}$ term we have used here that ${\bar \theta}^m\c_p\theta^n=0$ for electric gaugings as a consequence of the quadratic constraint.

We find that the complete variation of the topological term is given by
\bea
e^{-1}\delta \cL_{\rm top} &=&
-\ft1{8\s2}\,e^{-1}\,\epsilon^{\mu\nu\rho\sigma\kappa\lambda}{\bar\cH}_{\mu\nu}{} \c^M \cH_{\rho\sigma} \, (\Delta B_{\kappa\lambda\,M})
\label{varLtop}\w2
&& -\ft12 j\cH_m\,\cdot D (\Delta B^m) -\ft1{3\s2}\, g j\cH^m \cdot \, ({\bar\theta}_m \Delta C) -\ft1{\s2}\, j\cH_m\,\cdot ({\bar\cH} \c^m\,\Delta A)\ ,\nn
\eea
and thus expressible in a very compact form in terms of the covariant variations $\Delta$ defined above.
In the ungauged theory, only the first and the last term of this variation are present, while the second term becomes a total derivative. Note that the variation (\ref{varLtop}) is only ${\rm GL}(5)$ invariant. This forbids for example in the gauged theory to integrate by parts the second term, as the sum over $m$ is not the full ${\rm SO}(5,5)$ covariant one whereas the derivative $D_\sigma$ is covariant with respect to a gauge group that might not be contained in ${\rm GL}(5)$. Only together with the variation of the kinetic term
\bea
e^{-1}\delta \cL_{\rm kin}(A, B) &=& -\cH^{\mu\nu\,A}\,\cM_{AB}
\,\Big(D_{\mu}(\Delta A_\nu{}^B) -\ft1{\s2}\,g\theta^{BM}\,\Delta B_{\mu\nu\,M}\Big)
\\[.8ex]
&&{}
-\ft16 \cH_m\,\cdot K^{mn}\, \left( 3D(\Delta B_n)+3{\s2}\,{\bar\cH} \c_n \,\Delta A - {\s2}\,g {\bar\theta}_n\,\Delta C \right)\ ,
\nonumber\label{dL}
\eea
the two non-covariant terms join and the combined variation takes the form
\bea
e^{-1}( \delta\cL_{\rm kin}(A, B) + \cL_{\rm top}) &=& -\cH^{\mu\nu\,A}\,\cM_{AB}\,\Big(\cD_{\mu}(\Delta A_\nu{}^B)
- \ft1{\s2}\,g \theta^{BM}\,\Delta B_{\mu\nu\,M}\Big)
\nn\w2
&& -\ft1{8\s2}e^{-1}\,\epsilon^{\mu\nu\rho\sigma\kappa\lambda}
{\bar\cH}_{\mu\nu}{}\c^M \cH_{\rho\sigma}\,(\Delta B_{\kappa\lambda\,M})
\nn\w2
&&
-\ft12 j \cG_M \cdot \cD\Delta B^M -\ft1{\s2}\, j \cG_M\,\cdot ({\bar\cH} \c^M\,\Delta A)
\nn\w2
&& -\ft1{3\s2}\, g j (\cH_M-\cG_M)\cdot ({\bar\theta}^M\Delta C)\ .
\label{dkt}
\eea
Since there is no kinetic term for the $3$-form potential $C_A$, its bosonic field equation is given by
\be
g\theta_m^A \left(\cH^m-jK^{mn}\cH_n\right) =0\ ,
\label{ce}
\ee
where we have used \eq{d3}. As for the bosonic field equation of the ``magnetic'' 2-form potentials, it takes the form
\be
g\theta^A_m \left( \cH_{\mu\nu\rho\sigma\,A} + \ft12 e \epsilon_{\mu\nu\rho\sigma\kappa\lambda} M_{AB} \cH^{\kappa\lambda\,B}\right)=0\ ,
\ee
where we have used the Bianchi identity \eq{B2}. This equation, as expected, furnishes the duality relation between the three-form potentials and the vector fields.

The variation formula \eq{dkt} is also very useful in finding the gauge coupling constant dependent terms in the action and supersymmetry transformation rules that are needed for establishing supersymmetry. The supersymmetry variations, with undifferentiated supersymmetry parameter, that do not depend on the gauge coupling constant will be covariantizations of those which arise  in the ungauged Lagrangian. Therefore, they will cancel as in the ungauged theory, and in a covariantized form. Supersymmetric variations with overall explicit coupling constant dependence, on the other hand, cancel as follows:

(1)  The partial integration in the $\cG_M \cdot D \Delta B^M$ term yields  $g \cH_{\mu\nu\rho\sigma\, A}$ via the modified Bianchi identity \eq{B2}. This is canceled by a term arising in the variation of the gravitino in the Pauli coupling term $\cG\cdot \cO$, followed by partial integration, and use of the Bianchi identity \eq{B2}.

(2) The terms involving $g\cH^A$ coming from the $\cH \cdot \Delta B$ term in \eq{dkt} and the new variations of the Pauli term $J\cH$, are canceled by the terms coming from the old variations in the fermionic mass terms, in gravitino kinetic term and the Noether term $P_\mu I^\mu$, using \eq{4c} and \eq{4b}.

(3) The terms involving  $g \cH_M$ coming from the $\cH_M \cdot \Delta C$ term in \eq{dkt}, cancel the terms coming from the variation in the Pauli term $J \cH$ using the modified Bianchi identity \eq{B1}. In fact, this is a convenient way to determine the supersymmetric variation of $C_A$.

(4) The terms involving  $g\cG_M$ coming from the $\cG_M \cdot \Delta C$ term in \eq{dkt}, the new variations of the Pauli coupling term $\cG\cdot \cO$ and the old variations of the $g$-dependent fermionic mass terms, all cancel.

(5) Finally, the new variations of the fermionic mass terms and the old variations of the potential, all cancel.


\subsection{Classification of Gaugings Under ${\rm
GL}(5)$}\label{sec:discussion}


So far, we have shown that every tensor $\theta^{\cA M}$ in the ${\bf 144}_{c}$ of ${\rm SO}(5,5)$ which satisfies the quadratic constraint~(\ref{quad}) defines a consistent and maximally supersymmetric gauging in six dimensions. It remains to study the possible solutions of~(\ref{quad}) and to identify the resulting theories. As usual, a systematic way to scan the various possibilities is given by decomposing $\theta^{\cA M}$ under a given subgroup of ${\rm SO}(5,5)$ and to separately analyze the different irreducible parts. In six dimensions, a distinguished subgroup is the maximal ${\rm GL}(5)\subset {\rm SO}(5,5)$ which allows to identify a possible seven-dimensional origin of the theories --- with ${\rm SL}(5)$ corresponding to the seven-dimensional duality group --- as well as a possible origin in eleven dimensions, in which context ${\rm GL}(5)$ is associated to the five-torus on which the
reduction is performed.

Under ${\rm GL}(5)$, the ${\rm SO}(5,5)$ representations break as
\bea
{\bf 10}\rightarrow 5^{+2}+5'{}^{-2}\;,
\quad
{\bf 16}_{s}\rightarrow 1^{-5}+5'{}^{+3}+10{}^{-1}
\;,
\quad
{\bf 16}_{c}\rightarrow 1^{+5}+5^{-3}+10'{}^{+1}
\;,
\label{gl5}
\eea
where we denote the $B_m$ by $5$ and the $B^m$ by $5'$.
The adjoint breaks as
\bea
{\bf 45} \rightarrow 1^0 + 24^0 + 10^{+4} + 10'{}^{-4} \;.
\label{adjoint}
\eea
The $1^0 + 24^0$ is the ${\rm GL}(5)$ subgroup, the $10^{+4}$ generators
are realized as shift symmetries on the scalar fields.
They correspond to the off-diagonal block $z$ in (\ref{dt1}) and thus correspond
to
off-shell symmetries of the Lagrangian. The complete off-shell
symmetry group is thus given by ${\rm GL}(5)\ltimes 10^{+4}$.
The $10'{}^{-4}$ generators on the other hand are hidden symmetries that
correspond to the
off-diagonal block $y$ in (\ref{dt1}) and are realized
only on-shell, i.e.\ do not constitute symmetries of the action.
We expect that there is a dual Lagrangian in which the $10^{+4}$ and
$10'{}^{-4}$ generators
have exchanged their roles.

Next, we turn to the classification of gaugings under ${\rm GL}(5)$.  Under ${\rm
GL}(5)$, the embedding tensor ${\bf 144}_c$ decomposes as
\bea
{\bf 144}_c
&\rightarrow&
5'{}^{+3}+5^{+7}+10^{-1} +15^{-1} + 24^{-5} +40'{}^{-1} +45'{}^{+3}
\;.
\label{split144}
\eea
Splitting $\theta^{A M} = (\theta^{A m},\theta^A_m)$ this amounts to distinguishing between electric and magnetic gaugings: gaugings triggered by $\theta^{A m}$ only involve the electric two-forms $B_m$ and no three-forms. This can be seen explicitly in the tensor gauge transformations~(\ref{gauge}), the covariant field strengths~(\ref{covAH}) and the topological term~(\ref{Ltopelectric}). On the other hand, gaugings triggered by $\theta^A_m$ involve magnetic two-forms $B^m$ as well as additional three-form tensor fields. In terms of representations, these components can contain
\bea
\theta^{A m} =
5'{}^{+3}+10^{-1} + 24^{-5} +40'{}^{-1} \;,
\qquad
\theta^A_m =
5'{}^{+3}+5^{+7}+10^{-1} +15^{-1} +45'{}^{+3}
\;.
\label{splitem}
\eea
Comparing this to~(\ref{split144}), we see, that $24^{-5} +40'{}^{-1}$ and $5^{+7}+15^{-1} +45'{}^{+3}$ trigger purely electric and purely magnetic gaugings, respectively, whereas $5'{}^{+3}+10^{-1}$ correspond to gaugings involving simultaneously electric and magnetic two-forms. Recall the quadratic constraint
\bea
\theta^{A M}\,\theta^{A N}\,\eta_{MN}&=& 0\;,
\qquad
\theta^{A M}\,\theta^{B [N}\,
(\gamma^{P]})_{AB}~=~0\;.
\label{quadA}
\eea
The first equation is automatically satisfied for gaugings that are purely
electric or purely magnetic.
For these we have to impose only the second equation, which is a ${\bf 320}$
under ${\rm SO}(5,5)$
and thus
\bea
{\bf 320} &\rightarrow&
5^{+2}+5'{}^{-2}+40^{+6}+40'{}^{-6}
+45^{-2}+45'{}^{+2}+70^{+2}+70'{}^{-2} \;.
\label{320}
\eea

This shows that e.g.\ any $\theta$ in the $24^{-5}$ (since its square does not show up in (\ref{320})) defines a consistent (electric) gauging. In fact, this makes sense: these are the Scherk-Schwarz gaugings obtained by reduction from seven dimension, the $24^{-5}$ corresponds to choosing a generator in the seven-dimensional symmetry group ${\rm SL}(5)$. The $40'^{-1}$ on the other hand also defines purely electric gaugings, but these $\theta$'s need to satisfy an additional quadratic constraint in the $70'{}^{-2}$ of (\ref{320}). These are the theories obtained by torus reduction from gaugings in seven dimensions, where indeed (part of) the embedding tensor lives in the $40'$ and its quadratic constraint in the $70'$~\cite{Samtleben:2005bp}. Explicitly, for $\theta$ given by $\vartheta^{mn,k}=\vartheta^{[mn],k}$ with $\vartheta^{[mn,k]}=0$,
the quadratic constraint is
\bea
\vartheta^{mn,r}\vartheta^{pq,s}\,\epsilon_{mnpqk}&=&0\;.
\label{q70}
\eea

Purely magnetic gaugings described by the $5^{+7}$ also satisfy automatically the quadratic constraint (\ref{320}). They may correspond to reductions from eleven dimensions with non-trivial four-form flux. Also for magnetic gaugings described by the $15^{-1}$, the square of $\theta$ does not show up in (\ref{320}), thus these are automatically consistent theories. They come from torus reduction of seven-dimensional ${\rm CSO}(p,q,r)$ gaugings~\cite{Pernici:1984xx,Samtleben:2005bp}, whose embedding tensor indeed transforms in the $15$. And it makes perfect sense that these give magnetic gaugings: in order to gauge ${\rm CSO}(p,q,r)$ in seven dimensions, a number of two-forms have been dualized into three-forms, whose reduction to six dimensions gives rise to the magnetic dual two-forms. A more constrained version of magnetic gaugings is parametrized by the $45'{}^{+3}$ (explicitly: some traceless $\vartheta^{mn}_r=\vartheta^{[mn]}_r$) with a quadratic constraint in
the $40^{+6}$, given by
\bea
\vartheta^{mn}_r \,\vartheta^{pq}_{[s}\,\epsilon^{\phantom{[p]}}_{k]mnpq} &=& 0
\;.
\label{q40}
\eea
Note the duality of this constraint to~(\ref{q70}). As $\vartheta^{mn}_r$ has the index structure
of a torsion, these theories could
presumably be obtained by reduction from eleven dimensions on some twisted tori.

The gaugings triggered by $5'{}^{+3}$ and $10^{-1}$ (let us parametrize them by $\vartheta^{m}$ and $\vartheta_{mn}=\vartheta_{[mn]}$, respectively) are neither purely electric nor purely magnetic, i.e.\ the first equation of (\ref{quadA}) has to be imposed explicitly. However, it follows immediately that they give rise to only few constraints. While apparently they cannot be switched on together, $\vartheta^{m}$ alone defines a consistent gauging,
and $\vartheta_{[mn]}$ comes with the constraint
\bea
\vartheta_{kl}\vartheta_{mn}\,\epsilon^{klmnp} &=& 0\;,
\eea
which is solved by $\vartheta_{[mn]}=\lambda_{[m}\xi_{n]}$, which is a possible
candidate to be the most general solution.

Of course, there are many more gaugings possible which correspond to
simultaneously switching on various ${\rm GL}(5)$ irreducible components of $\theta$.

The nature of these gaugings can be illustrated by the following table
\begin{equation}
\begin{tabular}{c|cccc}
$\Theta_{A}{}^{MN}$&
$10'{}^{-4}$&$1^0$& $24^0$ & $10^{+4}$\\
\hline
$5^{-3}$&$5^{+7}$
&$5'{}^{+3}$
&$(5'+45')^{+3}$
&$(10+40')^{-1}$
\\
$10'{}^{+1}$
&$(5'+45')^{+3}$
&$10^{-1}$
&$(10+15+40')^{-1}$
&$24^{-5}$
\\
$1^{+5}$
& $10^{-1}$ &
&$24^{-5}$&
\\
\end{tabular}
\label{couplings}
\end{equation}
where the top row represents the ${\rm SO} (5,5)$ generators, the left column represents the vector fields, and we have depicted their mutual couplings by the various ${\rm GL}(5)$ components of the embedding tensor according to \eq{et}. In accordance with the discussion above, we see that electric gaugings (those triggered by the $24^{-5} +40'{}^{-1}$) involve only generators that belong to the off-shell symmetry group ${\rm GL}(5)\ltimes 10^{+4}$ of the Lagrangian. Magnetic gaugings in the $5^{+7}+45'{}^{+3}$ on the other hand also gauge symmetries that are realized only on-shell, very much like what happens in other even dimensions. A notable exception are gaugings triggered by the $15^{-1}$, these are magnetic in the sense that they require introduction of magnetic two-forms and three-form fields, on the other hand they only gauge on-shell symmetries inside of ${\rm GL}(5)$! This is rather different from the situation in four dimensions, where every gauging whose gauge group resides within the off-shell symmetry group of the Lagrangian can be realized as a purely electric gauging, i.e.\ without introduction of magnetic forms~\cite{deWit:2005ub}. Note however that due to the first quadratic constraint in~(\ref{quad}) there is always a frame, which may be reached by an ${\rm O}(5,5)$ rotation from Tanii's Lagrangian, in which the gauging takes a purely electric form. However, this may not be the frame the most suited in order to identify a particular higher dimensional origin.


\subsection{Classification of Gaugings Under ${\rm
SO}(4,4)$ and Truncation to $N=(1,1)$ Theories}
\label{sec:discussion2}


It would be interesting to consider truncations of our results to $D=6$ half-maximal gauged supergravity. The duality group of non-chiral $D=6$ half-maximal gauged supergravity coupled to $4+n$ vector multiplets is given by ${R}^+\times {\rm SO}(4,4+n)$. There are three different classes of gaugings~\cite{Bergshoeff:2007vb}. The gauging of the ${ R}^+$ scaling symmetry leads to an embedding tensor in the fundamental representation of the duality group. On the other hand, the gauging of a subgroup of the ${\rm SO}(4,4+n)$-factor leads to an embedding tensor in the three-index antisymmetric representation. On top of this there is also a massive supergravity with an embedding tensor in the fundamental representation. This includes the massive supergravity of~\cite{Romans:1985tw}. Gaugings of this theory coupled to further matter multiplets have been constructed in~\cite{D'Auria:2000ad,Andrianopoli:2001rs}. The IIA origin of the $n=16$ case via a $K3$ compactification was studied in \cite{Haack:2001iz}. A massive supergravity is a particular deformation of the $p$-form gauge transformations that does not involve the gauging of a duality group. These massive supergravities are also described by the embedding tensor approach. The T-duality properties of the $D=6$ half-maximal massive supergravities have been discussed in \cite{Janssen:2001hy,Behrndt:2001ab}.

Let us see, how these structures can be embedded into our results. The duality group of the half-maximal supergravity coupled to $4$ vector multiplets embedded in the maximal theory is ${R}^+\times {\rm SO}(4,4)$ under which the ${\rm SO}(5,5)$ representations break according to
\bea
{\bf 10} &\rightarrow&
8_{v}^{0}+1^{+2}+1^{-2}\;,
\qquad
{\bf 16}_{c} ~\rightarrow~
8_{c}^{+1}+8_{s}^{-1}
\;,
\nonumber\\
{\bf 45} &\rightarrow&
1^{0}+28^{0}+8_{v}^{+2}+8_{v}^{-2}
\;.
\eea
In particular, the embedding tensor breaks according to
\bea
{\bf 144}_{c} &\rightarrow&
56_{c}^{-1}+56_{s}^{+1}+
8_{c}^{-1}+8_{s}^{+1}+
8_{c}^{+3}+8_{s}^{-3}
\;,
\label{emb44}
\eea
and we may analyze the gaugings triggered by the different
${\rm SO}(4,4)$ irreducible parts.
The three different classes discussed above correspond to the gaugings induced by
the $8_{c}^{-1}$, the $56_{c}^{-1}$ and the $8_{c}^{+3}$, respectively.
Again, we can infer the structure of these gaugings from the table of minimal couplings
\begin{equation}
\begin{tabular}{c|cccc}
$\Theta_{A}{}^{MN}$
&$8_{v}^{-2}$
&$1^{0}$
&$28^{0}$
&$8_{v}^{+2}$
\\
\hline
$8_{c}^{+1}$
&$8^{+1}_{s}+56^{+1}_{s}$
&$8^{-1}_{c}$
&$8^{-1}_{c}+56^{-1}_{c}$
&$8_{s}^{-3}$
\\
$8_{s}^{-1}$
&$8_{c}^{+3}$
&$8^{+1}_{s}$
&$8^{+1}_{s}+56^{+1}_{s}$
& $8^{-1}_{c}+56^{-1}_{c}$
\\
\end{tabular}
\label{couplingsD4}
\end{equation}
where again the top row and the left column represent the ${\rm SO} (5,5)$ generators and the vector fields of the maximal theory, respectively, and we have depicted their mutual couplings by the various ${\rm SO}(4,4)$ components of the embedding tensor according to \eq{et}. The structure of the deformed $p$-form tensor hierarchy can be illustrated by explicitly branching the matrix $\theta^{AM}$
\begin{equation}
\begin{tabular}{c|ccc}
$\theta^{AM}$
&$1^{-2}$
&$8_{v}^{0}$
&$1^{+2}$
\\
\hline
$8_{c}^{+1}$
&$8^{+3}_{c}$
&$8^{+1}_{s}+56^{+1}_{s}$
&$8_{c}^{-1}$
\\
$8_{s}^{-1}$
&$8_{s}^{+1}$
&$8^{-1}_{c}+56^{-1}_{c}$
& $8^{-3}_{s}$
\\
\end{tabular}
\label{couplingsD41}
\end{equation}
which plays the role of the intertwiner between vectors/2-forms and 2-/3-forms, respectively, cf.~(\ref{covAH}). Truncation to the half-maximal theory coupled to $4$ vector multiplets corresponds to projecting out the $8_{s}^{-1}$ vector fields and the $8_{v}^{0}$ two-forms, in the bosonic sector. Next, we describe the two classes of gaugings of this theory triggered by the $8^{+3}_{c}$ and $8^{-1}_{c}$.

Let us first consider the gaugings induced by the $8^{+3}_{c}$. As its square does not appear in the decomposition of the quadratic constraint ${\bf 10}+{\bf 126}_{c}+{\bf 320}$, a gauging induced by such an embedding tensor $\vartheta^{\alpha}$ is automatically consistent. According to~(\ref{couplingsD4}), it gauges the $8_{v}$ shift symmetries, while~(\ref{couplingsD41}) shows that it induces a St\"uckelberg type coupling of the form ${\cal F}_{\mu\nu}{}^{\alpha}+\vartheta^{\alpha} B_{\mu\nu}$. Alternatively, we may consider the gaugings induced by the component $8^{-1}_{c}$ gaugings induced by the component $8_c^{-1}$ which we shall denote by
$\tilde{\vartheta}^\alpha$. As the quadratic constraint contains a $1^{-2}$, we
deduce that $\tilde{\vartheta}^\alpha$ should be a null vector
($\tilde{\vartheta}^\alpha \tilde{\vartheta}_\alpha =0$).
This defines another class of viable gaugings. According to~(\ref{couplingsD4}) these in particular gauge the $R^{+}$ shift symmetry. Note however, that $8^{+3}_{c}$ and $8^{-1}_{c}$ cannot be switched on simultaneously, but lead to a quadratic constraint of the form $\vartheta^{(\alpha}\tilde{\vartheta}^{\beta)}=0$. This is in line with the occurrence of corresponding $6$-form potentials in the same representations \cite{Bergshoeff:2007vb,dWS,dWS2}.

The $4$ vector multiplets in these theories can be consistently truncated to obtain the pure half-maximal theory~\cite{Romans:1985tw}. It is well known that there exists an $SU(2)$ gauged version of this theory with an additional massive deformation parameter. The $SU(2)$ gauge group is the non-chiral diagonal subgroup of the ${\rm SU}(2) \times {\rm SU}(2)$ isomorphism group of the $N=(1,1)$ Poincar\'e  superalgebra. It is interesting to determine if and how this theory can be embedded in the gauged maximal theory. To this end, considering the gaugings induced by the $8_c^{+3}$ discussed above, upon a consistent truncation to the pure half-maximal theory, the shift symmetries and the associated vector fields $8_{s}$ are projected out and what remains is precisely Romans' massive deformation. In this theory, the only effect of the gauging in the bosonic sector is the St\"uckelberg type coupling and the scalar potential, the mass parameter $m$ corresponding to a fixed component within $\vartheta^{\alpha}$. Thus, we are able to show how Romans' massive deformation of the pure half-maximal theory can be embedded into the maximal theory where it is a true gauging of shift isometries.

We can show that the ${\rm SU}(2)$ gauging with mass parameter set to zero follows from a suitable truncation as well. In fact, there exists a variant of Romans' theory \cite{Kerimo:2003am,Kerimo:2004md} emerging in a generalized  Kaluza-Klein reduction of D$=$11 supergravity on $K3\times R$, with all $4$ vectors abelian, which should also be embeddable in gauged maximal supergravities. However, it remains an open question if Romans' theory with non-vanishing gauge coupling constant and mass deformation parameter can be embedded in the maximal theory. In general, the lower supersymmetric $6D$ supergravities admit more general couplings than those which can be obtained by truncation of the maximal theory since the quadratic constraints encountered in gauging of the maximal theory are far more stringent than what is required in gauging of the lower supersymmetric theories. In fact, a very simple example of this phenomenon arises in seeking a truncation of Romans' theory to an $N=(1,0)$ supergravity that maintains any gauging at all. One quickly finds that this is not possible, and indeed this is the case for the variant of the Romans' theory as well. On the other hand, a $U(1)$ gauged $N=(1,0)$ supergravity does exist in its own right, and it is constructed directly in the $N=(1,0)$ supersymmetric setting \cite{Nishino:1984gk,Salam:1984cj}.

In conclusion, it would be highly interesting to see, which gaugings of the half-maximal theory, or indeed minimal theory, with or without matter couplings, can be lifted to the maximal gaugings and which of their solutions may be embedded. We leave these and related questions for future work.

\bigskip
\bigskip

\noindent
{\bf Acknowledgement}\\
\noindent

We are very grateful to Yoshiaki Tanii for helpful discussions.
Part of the calculations has been facilitated by use of the computer algebra
system Cadabra~\cite{Peeters:2006kp,Peeters:2007wn} and we thank Kasper Peeters for support.
E.S. would like to thank University of Groningen for hospitality where this project was conceived.
E.B.~is supported by the European Commission FP6 program MRTN-CT-2004-005104 and
by the Spanish grant BFM2003-01090. The work of H.S. is supported in part by the Agence
nationale de la recherche (ANR), and E.S. is supported in part by NSF Grant PHY-0555575.

\newpage

\begin{appendix}

\section{Notations and Conventions}

In our conventions:
\bea
\{\c_r,\c_s\} &=& 2\eta_{rs}\ ,\qquad \eta_{rs} ={\rm diag}(-,+,+,+.+,+)\ \nn\w2
\{\C_\unA,\C_\unB \} &=& 2\eta_{\unA\unB}\ ,\quad\  \eta_{\unA\unB} ={\rm diag}(+,+,+,+,+,-,-,-,-,-)\ ,
\eea
where $\unA=(a,\adot)$. Moreover, $\c_{r_1...r_6}=\epsilon_{r_1...r_6} \c_7$ and $(\c_7)^2=1$. A convenient representation for $\C_\unA$ is
\be
\C_a = 1\times \c_a \times \sigma_1\ ,\qquad \C_\adot = \c_\adot \times 1\times i\sigma_2 \ ,
\ee
with
\bea
\{\c_a,\c_b\} &=& 2 \delta_{ab}\ ,\qquad \delta_{ab}={\rm diag}(+,+,+,+,+)\ ,\nn\w2
\{\c_\adot,\c_\bdot\} &=& 2 \delta_{\adot\bdot}\ ,\qquad \delta_{\adot\bdot}={\rm diag}(+,+,+,+,+)\ ,
\eea
From the position where they are used, it can be seen that the matrix $\c^a$ is either $(\c^a)_\a{}^\b$ or
$(\c^a)_{\a\ad}{}^{\b\bd}= (\c^a)_\a{}^\b\,\delta_\ad^\bd$, depending on what it acts on, and similarly for $\c^\adot$. The indices $(a,\adot)$
on the $\c$-matrices are raised and lowered with $\delta_{ab}$ and $\delta_{\adot\bdot}$. We use the chirally
projected ${\rm SO}(5,1)$ Dirac matrices, such that $\c_\mu$ are symmetric and $\c_{\mu\nu\rho}$ are antisymmetric.
Similarly, we use the chirally projected ${\rm SO}(5,5)$ Dirac matrices and all (anti) symmetrizations are with unit
strength. Note that there is no need to raise and lower the spinor indices in this chiral notation. The $USp(4)$
indices are raised and lowered by the symplectic invariant tensors as: $X^\a=\Omega^{\a\b}X_\b$,
$X_\a=X^\b\Omega_{\b\a}$ with $\Omega_{\a\b}\Omega^{\b\c}=-\delta_\a^\c$. The symmetry properties of the $\c$ and
$\C$ matrices are as follows:
\bea
&& \c_\mu C:\ {\rm symmetric}\ \ , \qquad\ \  \c_{\mu\nu\rho}C:\ {\rm antisymmetric}\nn\w2
&& (\c_a)_{\a\b}:\ {\rm antisymmetric}\  , \qquad (\c_{ab})_{\a\b} :\ {\rm symmetric}\nn\w2
&&(\c_M, \c_{M_1\cdots m_5})_{AB}:\ {\rm symmetric}\ ,\quad (\c_{MNP})_{AB}:\ {\rm antisymmetric}
\eea
The ${\rm SO}(5)$ $\c$-matrices satisfy the identity
\be
(\c^a)_\a{}^\b (\c^a)_\c{}^\delta = 2 \delta_\a^\d \delta^\b_\c + 2\Omega_{\a\c}\Omega^{\b\d} -\delta_\a^\b\delta_\c^\d\ .
\label{cc5}
\ee
Note also that any $A_{\a\b}=-A_{\b\a}$, and any $S_{\a\b}=S_{\b\a}$ can be expanded as
\bea
A &=& \ft14 \tr\,A + \ft14  \c^a \tr\,\c^a A\ ,
\label{cc6}
\w2
S &=& -\ft18 \c^{ab}\tr\,\c^{ab} S\ .
\label{cc7}
\eea
The matrices $\cV^A_{\a\ad}$ and $\cV_{\cA\a\ad} (\cA=1,...,16,\ \ \a,\ad=1,...,4)$ can
be treated as sixteen $4\times 4$ matrices $V^A$ and $V_\cA$. The index $\cA$ is a chiral ${\rm SO}(5,5)$ spinor
index which is never raised and lowered but the $\a$ and $\ad$ indices can be raised and lowered as usual.

Whenever the row and column indices of a matrix are suppressed we will always assume that the indices are in the order $(M)_\star{}^\star$, with the exception of the chirally projected ${\rm SO}(5,1)$ Dirac matrices $\c_\mu$ and again chirally projected ${\rm SO}(5,5)$ matrices $\c^M$, in which case they are both up or down. Thus, for example,
\bea
{\bar\psi} \c^a \chi &=& {\bar \psi}^\a(\c^a)_\a{}^\b\,\chi_\b\ ,\quad {\bar\psi} \c^a V_\cA \chi = {\bar\psi}^\a (\c^a)_\a{}^\b\,(V_\cA)_\b{}^\bd \chi_\bd\ ,
\nn\w2
{\bar V}\c^M V &=& V^{\cA \a\ad} (\c^M)_{\cA\cB} V^\cB_{\a\ad} \ ,\quad {\bar V} \c_{MN} V= V^{\a\ad} (\c_{MN})_A{}^B V_{B \a\ad}\ .
\eea
Furthermore, ${\bar V}$ always denotes $V^{A\a\ad}$. Finally, our conventions for differential forms are as follows:
\be
\omega= \frac1{p!}\,dx^{\nu_1}\wedge \cdots dx^{\nu_p} \,\omega_{\nu_1...\nu_p}\ ,
\qquad dx^{\nu_1}\wedge\cdots dx^{\nu_6} = - e^{-1}\epsilon^{\nu_1...\nu_6} d^6 x \ .
\ee

\section{Useful Identities}

Proving invariance of the topological term (\ref{Ltop})
under tensor gauge transformations and showing that
its variation takes the fully covariant form (\ref{varLtop})
is quite lengthy and requires a number of rather
non-trivial identities which combine $SO(5,5)$ properties
with the constraints on the embedding tensor $\theta^{AM}$.
Among the ${\rm SO}(5,5)$ identities are
\bea
0&=& \gamma_{M\,A(B}\,\gamma^{M}{}_{CD)}\;,
\nonumber\\[1ex]
0&=&\gamma_{K\,A(C}\gamma^{MNK}{}_{D)B}-
\gamma_{K\,B(C}\gamma^{MNK}{}_{D)A}+
\gamma_{K\,CD}\gamma^{MNK}{}_{AB}
+4 \gamma^{[M}{}_{A(C}\gamma^{N]}{}_{D)B}
\;.
\nonumber\\
\eea
The following identity holds upon antisymmetrization in indices $[ABC]$:
\bea
0 &=&
\gamma_{K\,AD}\gamma_{L\,EF}\gamma^{MKL}{}_{BC}
+2\gamma_{K\,AD}\gamma_{L\,B(E}\gamma^{MKL}{}_{F)C}
+ 4\gamma^M{}_{A(E} \gamma^{K}{}_{F)B}\gamma_{K\,CD} \;.
\eea
Another ${\rm SO}(5,5)$ identity (upon antisymmetrization in indices $[ABCD]$)
is given by:
\bea
0 &=&
10\,\gamma_{K\,AE}\gamma^{K}{}_{BF}\gamma^{PQM}{}_{CD}
+8\,\gamma^{(M}{}_{AE}\gamma^{K)}{}_{BF}\gamma^{PQ}{}_{K\,CD}
+10\,\gamma^Q{}_{AE}\gamma_{K\,BF}\gamma^{PMK}{}_{CD}
\nonumber\\[1ex]
&&{}
-10\,\gamma_{K\,AE}\gamma^{P}{}_{BF}\gamma^{QMK}{}_{CD}
-4\,\gamma^Q{}_{AF}\gamma_{K\,BE}\gamma^{PMK}{}_{CD}
+4\,\gamma_{K\,AF}\gamma^{P}{}_{BE}\gamma^{QMK}{}_{CD}
\nonumber\\[1ex]
&&{}
+2\eta^{PQ}\gamma_{K\,AE}\gamma_{L\,BF} \gamma^{MKL}{}_{CD}
-2\gamma_{K\,EF} \gamma^{KL(P}{}_{AB}\gamma_L{}^{Q)M}{}_{CD}
\nonumber\\[1ex]
&&{}
-2 \gamma^{PNK}{}_{AE}\gamma^{QL}{}_{N\,BF}\gamma^{M}{}_{KL\,CD}
-\gamma^{KN[P}{}_{EF}\gamma_{NKL\,AB}\gamma^{Q]ML}{}_{CD}
\;.
\eea
We derive this identity by first observing that there must be a relation between this number of terms with this symmetry structure in the free indices, as a consequence of representation theory. We then compute the coefficients either
by tracing or by using an explicit representation.

Now we multiply this identity with $\theta_{P}^{E}\theta_{Q}^{F}$ and use
the tracelessness $\gamma^M{}_{AB}\,\theta_{M}^{B}=0$, upon which this identity
reduces to
\bea
0 &=&
\theta_{P}^{E}\theta_{Q}^{F} \;\;\Big(
3\gamma_{K\,AE}\gamma^{K}{}_{BF}\gamma^{PQM}{}_{CD}
+2\gamma^{M}{}_{AE}\gamma^{K}{}_{BF}\gamma^{PQ}{}_{K\,CD}
-6\gamma^Q{}_{AE}\gamma_{K\,BF}\gamma^{PMK}{}_{CD}
\nonumber\\[1ex]
&&{}\qquad \qquad
+\eta^{PQ}\gamma_{K\,AE}\gamma_{L\,BF} \gamma^{MKL}{}_{CD}
+\gamma_{K\,EF} \gamma^{KLP}{}_{AB}\gamma_L{}^{QM}{}_{CD} \;\Big)
\;.
\eea
Finally we may use the quadratic constraint on $\theta$ and obtain
\bea
0 &=&
\theta_{P}^{E}\theta_{Q}^{F} \;\Big(
3\gamma_{K\,AE}\gamma^{K}{}_{BF}\gamma^{PQM}{}_{CD}
+2\gamma^{M}{}_{AE}\gamma^{K}{}_{BF}\gamma^{PQ}{}_{K\,CD}
-6\gamma^Q{}_{AE}\gamma_{K\,BF}\gamma^{PMK}{}_{CD} \;\Big)
\;,
\nonumber\\
\eea
a quite strong identity (upon antisymmetrization in indices $[ABCD]$),
which enters the calculation of the variation of the topological term.

\end{appendix}

\pagebreak


\end{document}